\documentclass[conference]{IEEEtran}
\IEEEoverridecommandlockouts

\usepackage{cite}
\usepackage{amsmath,amssymb,amsfonts}
\usepackage{graphicx}
\usepackage{textcomp}
\usepackage{xcolor}
\usepackage{xspace}
\usepackage{comment}
\usepackage{kotex}
\usepackage{graphicx}
\usepackage{times}
\usepackage{latexsym}
\usepackage{tabularx}
\usepackage{multirow, lipsum, makecell}
\usepackage{amsmath}
\usepackage{algorithm}
\usepackage{algpseudocode}
\usepackage{booktabs}
\usepackage{caption}
\usepackage{subcaption}
\usepackage{enumitem}
\usepackage{color}
\usepackage{threeparttable}
\usepackage{tikz}
\usepackage{url}
\usepackage{xurl} 
\usepackage[hidelinks]{hyperref}

\definecolor{alizarin}{rgb}{0.82, 0.1, 0.26}
\definecolor{blue(ncs)}{rgb}{0.0, 0.53, 0.74}
\newcommand*\circledred[1]{\tikz[baseline=(char.base)]{%
            \node[shape=circle,text=white,fill=alizarin,inner sep=0.5pt] (char) {#1};}}
\newcommand*\circledblue[1]{\tikz[baseline=(char.base)]{%
            \node[shape=circle,text=white,fill=blue(ncs),inner sep=0.5pt] (char) {#1};}}

\def\BibTeX{{\rm B\kern-.05em{\sc i\kern-.025em b}\kern-.08em
    T\kern-.1667em\lower.7ex\hbox{E}\kern-.125emX}}

\makeatletter
\def\ps@IEEEtitlepagestyle{%
  \def\@oddfoot{%
    \hspace{1em}%
    \raisebox{3.2ex}{\normalsize\bfseries Accepted for publication in ICDE 2026}%
    \hfill
  }%
  \def\@evenfoot{}%
}
\makeatother

\begin{document}

\title{MatKV: Trading Compute for Flash Storage in LLM Inference\\}

\makeatletter
\newcommand{\linebreakand}{%
  \end{@IEEEauthorhalign}
  \hfill\mbox{}\par
  \mbox{}\hfill\begin{@IEEEauthorhalign}
}
\makeatother

\author{
Kun-Woo Shin\textsuperscript{\dag}, 
Jay H. Park\textsuperscript{\ddag}, 
Moonwook Oh\textsuperscript{\ddag}, 
Yohan Jo\textsuperscript{\dag}, 
Jaeyoung Do\textsuperscript{\dag}, 
Sang-Won Lee\textsuperscript{\dag} \\
\textsuperscript{\dag}Seoul National University, 
\textsuperscript{\ddag}Samsung Electronics \\
kunwooshin@snu.ac.kr, \{tino.park, mw.oh\}@samsung.com, \{yohan.jo, jaeyoung.do, swlee69\}@snu.ac.kr
}

\newcommand{\MATKV}{{MatKV}\xspace}
\maketitle

\newcommand{\module}[1]{\vspace{1ex}{\bf {#1}}\:\:}

\begin{abstract}
We observe two major trends in LLM-based generative AI: (1) inference is becoming the dominant factor in terms of cost and power consumption, surpassing training, and (2) retrieval augmented generation (RAG) is becoming prevalent. When processing long inputs in RAG, the prefill phase of computing the key-value vectors of input text is energy-intensive and time-consuming even with high-end GPUs. Thus, it is crucial to make the prefill phase in RAG inference efficient.
To address this issue, we propose \MATKV, a scheme that precomputes the key-value vectors (KVs) of RAG objects (e.g., documents), materializes them in inexpensive but fast and power-efficient flash storage, and reuses them at inference time instead of recomputing the KVs using costly and power-inefficient GPU.\footnote{https://github.com/kunwooshin/MatKV}  
Experimental results using Hugging Face’s Transformers library across state-of-the-art GPUs and flash memory SSDs confirm that, compared to full KV computation on GPUs, \MATKV reduces both inference time and power consumption by half for RAG workloads, without severely impacting accuracy in the question-answering task. 
Furthermore, we demonstrate that \MATKV enables additional optimizations in two ways. First, a GPU can decode text while simultaneously loading the materialized KVs for the next instance, reducing load latency. Second, since decoding speed is less sensitive to GPU performance than KV computation, low-end GPUs can be leveraged for decoding without significantly compromising speed once the materialized KVs are loaded into GPU memory.
These findings underscore \MATKV's potential to make large-scale generative AI applications more cost-effective, power-efficient, and accessible across a wider range of tasks and hardware environments.
\end{abstract}


\section{Introduction}

\module{LLM Inference} As generative AI (genAI) applications become increasingly ubiquitous, LLM (Large Language Model) inference speed has emerged as a critical factor in determining overall application latency. While much research has focused on optimizing LLM training, optimizing LLM inference is even more crucial due to its significantly larger market size and computational demand~\cite{PagedAtt, RAGCache}. LLM inference is inherently compute-intensive and power-hungry~\cite{Sarathi}, making efficient inference execution a key challenge in deploying large-scale genAI applications.

\module{Retrieval Augmented Generation} {Among various genAI workloads, retrieval-augmented generation (RAG) has become a dominant paradigm
because of its advantages, such as mitigating hallucinations, improving recency, and enhancing security~\cite{RAG}. In RAG, a user query is augmented with retrieved objects (e.g., documents), enabling the LLM to generate more knowledge-intensive and truthful answers compared to relying solely on its parametric knowledge. 

{However, RAG poses unique computational challenges~\cite{InfServing}.} 
In a typical RAG workflow, each retrieved chunk is usually quite large (e.g., 1,000 tokens per chunk)~\cite{llamaindex-chunksize}. Consequently, when LLM inference processes these chunks as part of the prompt, the prefill phase (i.e., computing the key-value vectors of the input chunks) becomes highly compute-intensive and power-demanding~\cite{Sarathi}. 
Moreover, many objects are repeatedly retrieved and prefilled across multiple queries, exacerbating the computational overhead. 
Considering the ever-growing popularity of RAGs and accordingly the exploding compute demand for LLM inference, optimizing inference efficiency in RAG environments is more critical than ever. 

\begin{figure}[t]
  \includegraphics[width=0.8\linewidth]{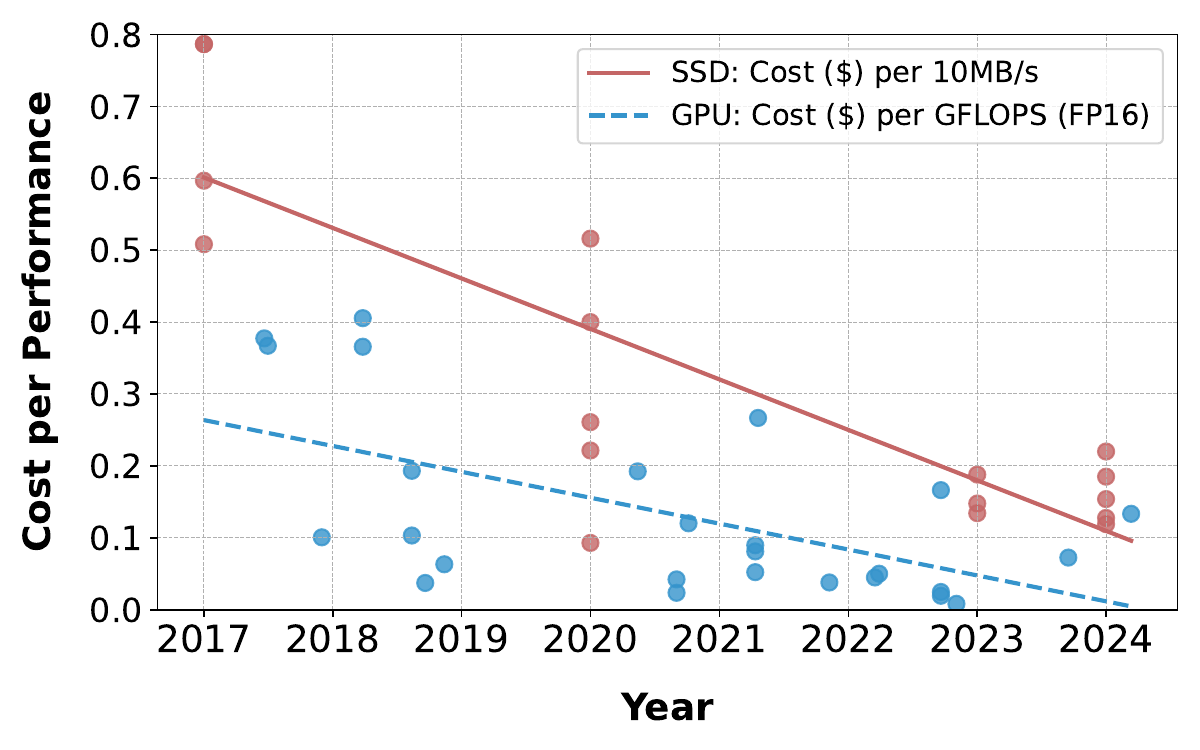}
  \centering
  \caption{GPU and SSD Cost/Performance Trend (2017 to 2024)}
  \label{fig:cost-per-perf}
  \vspace{-3.4mm}
\end{figure}

\module{Flash Memory SSDs} To meet the ever-growing compute demands in LLM training and inference, the compute power of modern GPUs has improved in terms of FLOPS and power efficiency. Likewise, flash memory SSDs have relentlessly improved in terms of cost, read bandwidth, and power efficiency, achieving astonishing performance metrics (Figure~\ref{fig:cost-per-perf}). 
For instance, Samsung's 9100 pro~\cite{9100Pro}, a commodity SSD, costs only \$0.1/GB but provides a read bandwidth of 14GB/second, and its active power consumption is only 7 watts. 

\module{MatKV (Materialized KVs)} In this paper, we propose \MATKV, a novel scheme designed to improve the efficiency of LLM inference in RAG workloads in terms of speed, power, and cost. It precomputes, when RAG objects are inserted to vector databases, their key-values once using GPU, materialize them on inexpensive but fast flash storage, and, whenever objects are retrieved to answer the given RAG query, reuse their corresponding KVs many times by loading them from flash storage to GPU HBM (High Bandwidth Memory), avoiding the full KV compute for RAG objects. That is, \MATKV trades the GPU compute for prefilling large RAG inputs with SSD storage for materializing their KVs.

The idea of \MATKV is motivated by three observations. 
First, RAG workloads frequently access long input chunks repetitively and thus lead to high computational demands during the prefill stage. Pre-computing the KVs for each chunk and retrieving them at inference time can significantly reduce both time and energy consumption. 
{Second, in RAG workloads that typically retrieve multiple objects, self-attention between retrieved chunks is generally less important (if not entirely unnecessary) compared to self-attention between the query and retrieved chunks. 
Third, considering the state of the art and trends in technological advancements in hardware, such as the performance, price, and power of GPUs and flash memory SSDs, trading GPU compute for storage to avoid the prefill stage is competitive and promising.} 

In this paper, we will answer three key questions. First, does \MATKV perform comparably to full KV compute using high-end GPU in terms of inference speed and power consumption?, Second, does \MATKV make economical sense? That is, is it affordable to materialize RAG objects and keep their KVs in flash storage? Third, given the technological advance trend in GPU and flash storage, is it promising in the future? For the last question, given that the \$/GB and bandwidth advances in flash storage technology outpace the FLOPs in GPUs, the answer is positive. To answer the second question, according to the formula used to derive the five-minute rule~\cite{fiveminrule}, we derive {\em the ten-day rule} stating that, for RAG objects which are retrieved every 10 days or more frequently, \MATKV is more economical than to compute KV using GPUs. 

To answer the first question, we prototyped MatKV by extending the popular Hugging Face Transformers library~\cite{HuggingFace-Transformer} with fewer than 500 lines of code changes, and conducted a comprehensive set of experiments using RAG datasets~\cite{TurboRAG, LongBench} with Nvidia H100 and RTX 4090 GPUs, Samsung 9100 Pro and PM9A3 SSDs, and three LLaMA models. The results show that MatKV outperforms full KV computation by up to 2× while reducing total system-wide power consumption by half. Beyond these gains, \MATKV also enables two novel system-level optimizations because of its decoupled prefill and decode phases. First, the GPU compute for decoding for a given query can overlap with SSD I/O for loading KVs for the next query, hiding KV loading latency.
Second, this decoupling enables inference on low-end GPUs: once KVs are materialized and retrievable via fast flash storage, even commodity GPUs or CPUs can handle large RAG inputs with performance comparable to high-end GPUs during prefill, broadening the deployment landscape for LLM inference.

Despite recent studies exploring KV cache reuse in LLM inference~\cite{RAGCache,CacheBlend,TurboRAG, DeepSeek}, prior works largely assume in-memory (DRAM-based) storage and focus on cache hit optimization without addressing the feasibility, performance, and cost implications of leveraging external storage. In contrast, MatKV demonstrates that materializing and retrieving precomputed key-value pairs from flash SSDs not only reduces inference latency and power consumption but also achieves superior cost-effectiveness compared to full GPU-based KV computation. Furthermore, the ten-day rule provides a principled and actionable guideline for when SSD-backed KV reuse becomes more economical than recomputation. 
Together, these empirical findings and analytical insights position MatKV as a practical and forward-looking solution for efficient LLM inference, especially in retrieval-heavy workloads such as RAG, and open a new direction for storage-centric optimization in large-scale generative AI systems.
\section{Background and Motivation\label{sec:bg}}

\subsection{LLM Inference and KV Cache}  

Large Language Model (LLM) inference is a computationally demanding process that requires substantial resources in terms of compute, memory bandwidth, and energy consumption~\cite{InfSurvey}. It consists of two primary phases: prefill and decoding. During the prefill phase, input tokens pass through multiple transformer layers, where the model computes the Key and Value representations for each token. In the decoding phase, the model generates new output tokens by autoregressively utilizing the precomputed KV tensors. In conventional LLM inference pipelines, KV tensors are recomputed on-the-fly for each retrieval instance, requiring significant computational resources from GPUs or other high-performance accelerators. This repeated computation introduces a major bottleneck, especially in Retrieval-Augmented Generation (RAG) scenarios, where retrieved objects often contain thousands of tokens. Consequently, the time required for the prefill phase increases super-linearly with input length, making it the dominant factor in inference latency~\cite{xiong2024layerkvoptimizinglargelanguage}. Even when using GPUs, the prefill phase alone can take tens to hundreds of milliseconds~\cite{CacheBlend}. Furthermore, the computational overhead of repeatedly generating KV tensors not only increases response time but also leads to excessive energy consumption.

To mitigate the overhead associated with repeatedly computing KV tensors, modern LLM inference pipelines employ KV caching~\cite{KVCaching}, where precomputed Key and Value representations are stored in memory. By caching these KV tensors, the model can bypass redundant computations, significantly improving efficiency during decoding. KV caching is particularly beneficial in scenarios where retrieved objects contain a large number of tokens, as it prevents the need to recompute attention keys and values for previously processed tokens. However, while KV caching reduces computational redundancy, it introduces a substantial memory footprint and data transfer. Since each token in a transformer-based LLM contributes to the KV cache, the total memory requirement grows proportionally with the number of tokens processed. In RAG applications, where retrieved objects can easily contain thousands of tokens per query, the KV cache size can quickly grow to several GBs for a single sequence, leading to 60\%-80\% of memory being wasted due to fragmentation and over-allocation~\cite{PagedAtt}.

Given the prohibitive memory requirements of maintaining large KV caches in memory, in this paper we argue that an effective alternative is to materialize KV vectors in high-performance storage rather than recomputing them on demand or caching them in memory. By storing precomputed KV tensors on storage devices such as SSDs, these values can be efficiently and economically retrieved when needed, enabling a trade-off between compute and storage. This approach not only alleviates GPU memory constraints but also significantly reduces inference latency and power consumption by avoiding redundant KV computation. 

\subsection{RAG Inference and Computational Challenges\label{sec:raginfandcomp}}

\begin{table}[tb]
\centering 
\caption{Average Number of Tokens in RAG Workloads}
 \label{tab:tokens}
\begin{tabular}{crr} \toprule
{Dataset}  & {Query}  & {Answer} \\
    \midrule 
CRAG~\cite{crag}& 15.56 & 11.17 \\
Trivia QA~\cite{triviaqa} & 18.16 & 4.05 \\
Google NQ~\cite{googlenq} & 10.09 & 5.77 \\
HotpotQA NQ~\cite{hotpotqa} & 23.11 & 3.53 \\
 \bottomrule
\end{tabular}
  \vspace{-3.4mm}
\end{table}

RAG has emerged as a dominant paradigm in generative AI, addressing limitations such as hallucination, knowledge recency, and security concerns by incorporating external knowledge sources~\cite{RAG}. In RAG, a vector database is an essential component as a knowledge source, which stores and indexes vectors created from documents using embedding models and then retrieves semantically relevant documents for the given user queries~\cite{chromadb}. For a defining characteristic of RAG is the large size of input prompts, which typically consist of multiple retrieved chunks, each containing hundreds or thousands of tokens. 
Unlike vanilla LLM inference scenarios, where input sequences are relatively short, RAG workflows involve extensive token processing before generation begins. Empirical analysis of common RAG datasets indicates that retrieved document chunks often contain an order of magnitude more tokens than the user query and output combined~\footnote{As summarized in Table~\ref{tab:tokens}, queries and generated responses in RAG workloads are relatively short - typically fewer than 20 tokens.}. For instance, in recent RAG implementations, default chunk sizes typically range from 800~\cite{openai-chunksize} to 1,024 tokens~\cite{llamaindex-chunksize} per document. Given that multiple chunks are often retrieved per query, the total input size can easily exceed several thousand tokens.

This disproportionately large input burden renders the prefill phase the dominant computational bottleneck in RAG inference. As the model must compute Key-Value tensors for all input tokens before decoding can commence, the prefill step accounts for the majority of inference time and energy consumption (for more details, see Section ~\ref{sec:basic-perf}). Given that RAG applications typically retrieve multiple such chunks, the aggregate computational cost escalates rapidly, making real-time inference infeasible in resource-constrained environments. Furthermore, RAG workloads exhibit high redundancy in retrieved document processing~\cite{RAGCache}. Retrieved chunks are frequently reused across multiple queries, yet existing inference pipelines redundantly recompute their KV caches for each inference request. This results in substantial inefficiencies in both computation and memory utilization. 

\subsection{Trading Compute for Storage for KV Cache in RAG\label{sec:trading-com-for-sto}}
The computational inefficiencies of RAG inference, particularly the redundant recomputation of KV caches for retrieved objects, present a fundamental challenge in optimizing both performance and energy consumption. \MATKV aims to mitigate this issue by precomputing KV caches and storing them in commodity flash storage, thereby eliminating the need for costly on-the-fly prefill computation using GPUs. This design choice not only reduces inference latency but also significantly lowers power consumption, making it a compelling alternative to conventional GPU-centric inference methods. 
 
To fully assess the feasibility and impact of \MATKV, three fundamental questions must be considered. First, can \MATKV match or even surpass GPU-based computation in terms of efficiency and performance? Given the high computational cost of the prefill phase, it is essential to evaluate whether retrieval from storage can serve as a viable replacement for real-time GPU processing. Second, does \MATKV offer an economic advantage over traditional inference methods? With the cost of GPUs remaining high and the price-performance ratio of SSDs improving, determining the economic trade-offs between compute and storage is crucial. Lastly, as model sizes continue to grow and hardware evolves, will \MATKV remain a scalable and promising solution in the long term? Understanding how future advancements in GPU processing power, SSD bandwidth, and their price changes will influence the feasibility of \MATKV is essential for assessing its longevity as an optimization strategy.

\module{\MATKV vs. GPU Compute} Consider a typical RAG inference scenario where a query retrieves a document chunk containing 1,024 tokens using a vector database. When computed on demand using an Nvidia H100 GPU~\cite{NVIDIA2}, prefilling 1,024 tokens for the retrieved chunk using LLaMA 3.1~\cite{llama} 70B requires approximately 500 milliseconds, consuming about 175 joules (with a peak power consumption of 350W), and the resulting KV cache occupies 250MB. Given the cost of an H100 GPU at \$50,000, this results in a significant computational burden, particularly when retrieval frequency is high. By contrast, a commodity SSD (e.g., the Samsung 9100 Pro SSD~\cite{9100Pro}), which costs \$400 per 4TB, can read the same 250MB KV cache in under 20 milliseconds while consuming less than 0.05 joules (with active power consumption of 7W), making it orders of magnitude more efficient than GPU-based computation.  
 
\module{Ten-Day Rule} The economic feasibility of this trade-off can be evaluated through a break-even analysis inspired by Jim Gray's five-minute rule~\cite{fiveminrule}. The break-even interval $T$, representing the minimum frequency at which a KV cache should be accessed to justify SSD storage over GPU recomputation, is given by:

\begin{equation}
T = \frac{\text{\$/GPU} \times \text{Sec/MB}} {\text{KVSize/GPU\_Sec} \times \text{\$/MB}}
\end{equation}
where \$/GPU, Sec/MB, KVSize/GPU\_Sec (MB unit), and \$/MB refer to the GPU price, the time taken to load 1MB from storage, the KV size generated by GPU per second, and the storage price (MB unit), respectively.
\\

Our analysis, based on the H100 GPU~\cite{NVIDIA2} and the commodity SSD~\cite{9100Pro} configuration described above, indicates that storing KV caches in SSDs is more cost-effective than GPU recomputation if a given document is accessed at least once every 10 days
\footnote{The above trade-off indicates that materializing KV caches for documents accessed at intervals of 10 days or longer is not cost-effective. However, for the sake of simplicity, this paper assumes that KV caches for all objects are materialized (see Section~\ref{sec:design} for more details). We leave the exploration of selective materialization as future work.}. Notably, when a document containing 1,024 tokens is retrieved once per hour, \MATKV is 100× more cost-efficient and achieves 2× lower latency compared to recomputation, as discussed in Section~\ref{sec:basic-perf}. With the growing adoption of large-scale RAG deployments, where frequently retrieved documents are accessed at much higher rates, \MATKV offers a significant advantage.

\begin{figure}[h]
  \includegraphics[width=\linewidth]{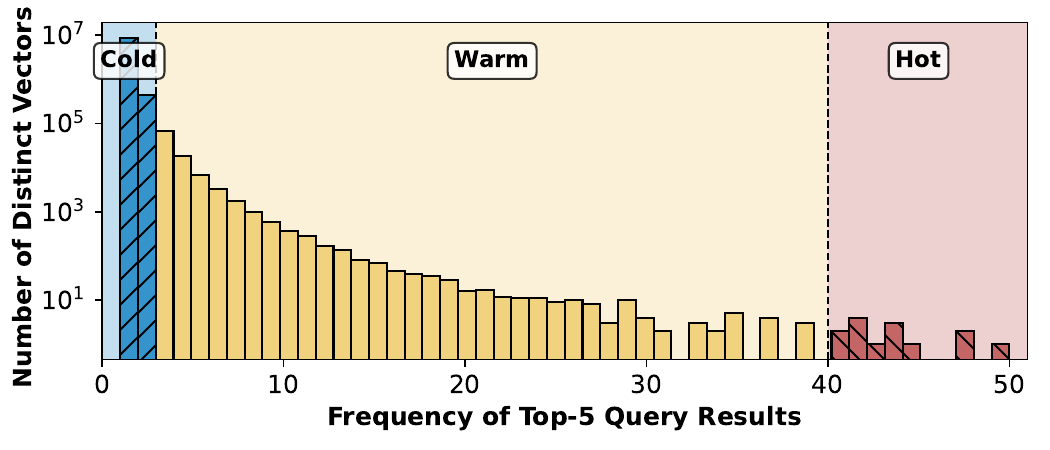}
  \centering
  \caption{Distribution of Accessed Vectors in RAG.
 While running 1M top-10 queries against a vector database with 9M document chunks, we measure the access frequency of each distinct vector.
 More than 900K documents are accessed twice or more. This skewed access indicates that there exists a large number of documents satisfying the ten-day rule.}
  \label{fig::skewedaccess}
\end{figure}

To illustrate the document access patterns in realistic RAG workloads, we measured the access frequency of each distinct document chunk while running 1M top-10 queries against the vector database of 9M documents from the deep1B benchmark~\cite{deep1b}, and plotted the result in Figure~\ref{fig::skewedaccess}. More than 900K documents (i.e., approximately 10\% of whole chunks) are accessed twice or more across different queries. This skewed access pattern is consistent with another observation~\cite{RAGCache} stating that a small fraction of documents accounts for the majority of retrieval requests.

\module{Effect of Technical Advances on \MATKV} Moreover, broader technological trends reinforce the long-term viability of \MATKV, as illustrated in Figure~\ref{fig:cost-per-perf}. Over the past decade, GPU FLOPS per dollar have improved approximately 10× every seven years, while SSD bandwidth has increased nearly 30×, with storage costs per gigabyte declining by an order of magnitude. These trends suggest that the economic and performance gap between compute and storage will continue to widen, making storage-based optimizations increasingly advantageous for future large-scale language model deployments.

In RAG scenarios where a single retrieved document’s KV cache can range from 100MB (e.g., LLaMA 3B with 1,000 tokens) to 1GB (e.g., LLaMA 70B), serving tens or hundreds of thousands of such documents would require several tens or hundreds terabytes of capacity. To materialize such a huge size of KV caches on DRAM is not economical in terms of cost and power consumption. Instead, we opt for flash storage as the scalable storage medium for materialized KV caches.
\section{Design of \MATKV\label{sec:design}}
\begin{figure*}[t]
\centering
  \begin{subfigure}[b]{0.40\textwidth}
    \includegraphics[width=\textwidth]{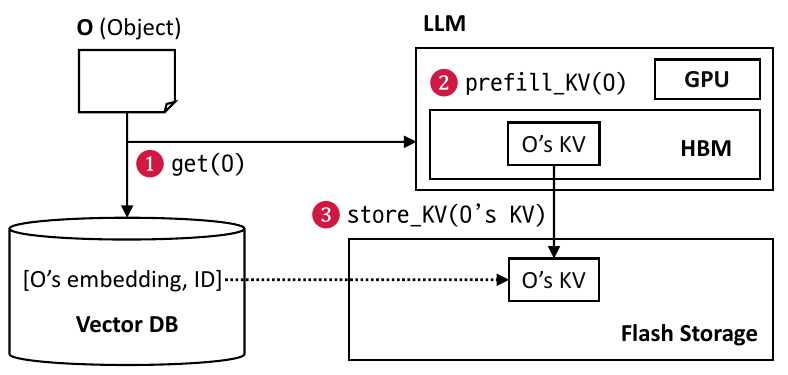}
    \caption{Materializing KV Caches for RAG Objects}
    \label{fig:get-store}
  \end{subfigure}
  \hspace{10mm}
  \begin{subfigure}[b]{0.40\textwidth}
    \includegraphics[width=\textwidth]{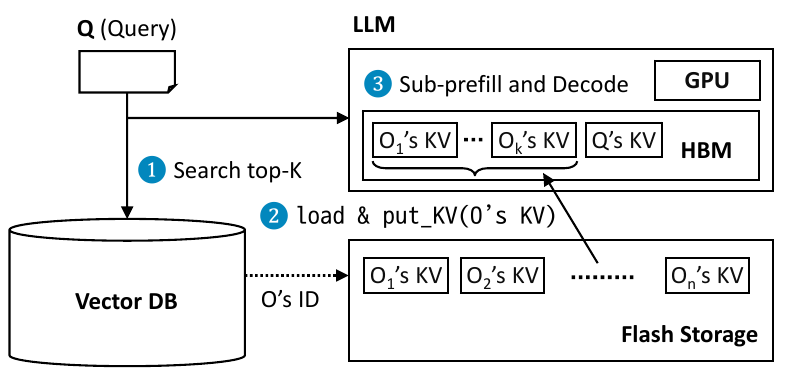}
    \caption{Efficient Retrieval of Precomputed KV}
    \label{fig:load-put}
  \end{subfigure}
  \caption{MatKV Architecture. (a) When a document (O) is inserted into the vector database, its KV cache is precomputed during the prefill phase using the LLM and stored in a commodity flash storage, enabling future reuse, (b) During inference, the system retrieves relevant documents based on the query (Q), loads their precomputed KV caches from flash storage, and directly utilizes them in the LLM, avoiding redundant prefill computations and improving inference efficiency.}
  \label{fig:matkv-archi}
\end{figure*}

\subsection{MatKV's Impact on Answer Quality}

\begin{table*}[h]
    \caption{QA Examples from 2WikiMultihopQA (LongBench)~\cite{2wikimqa, LongBench}}
    \label{tab:my_label}
    \centering
    \begin{tabularx}{\textwidth}{>{\arraybackslash}p{0.30\textwidth}|>{\arraybackslash}p{0.20\textwidth}|X|X}
    \toprule
        Question & Vanilla & MatKV & Answer \\
\midrule
Where was the wife of Francis I Rákóczi born?
& Ozalj, present day Croatia. & Ozalj, Croatia. & Ozalj \\ 
\midrule
Where did Helena Carroll's father study?
& St Patrick's College, Dublin. & St. Patrick's College, Dublin. & St Patrick's College \\ 
\midrule
Which film has the director died later, Seven In The Sun or Daughter Of The Jungle?
& Daughter Of The Jungle. & Daughter Of The Jungle. & Seven In The Sun \\ 
\bottomrule
    \end{tabularx}
    \label{tab:matkv-prelim}
    \vspace{-3mm}
\end{table*}

\MATKV is designed to improve LLM inference efficiency, particularly in Retrieval-Augmented Generation (RAG) workloads, by leveraging commodity flash storage to offload the computational burden of the prefill phase. Instead of repeatedly computing key-value (KV) caches for each retrieved document, as illustrated in Figure~\ref{fig:matkv-archi}, \MATKV precomputes these caches when documents are inserted into vector databases and stores them in flash SSDs\footnote{While MatKV assumes immediate KV cache materialization upon document insertion, an alternative is lazy materialization, where KV caches are generated only upon first retrieval. This reduces upfront computation but adds latency and complexity in caching. As MatKV prioritizes low-latency inference and high throughput, this study adopts immediate materialization.}. When an object is retrieved during inference, its KV cache is directly loaded from storage to GPU memory, eliminating redundant computations and reducing inference time, memory footprint, and power consumption.

\subsection{Materializing and Retrieving KV Caches}
At the core architecture of \MATKV is the materialization of KV caches in storage. Each document processed in a RAG pipeline undergoes a prefill phase, where its KV cache is computed and stored alongside the document in a vector database, as shown in Figure~\ref{fig:matkv-archi}a. \circledred{1} When a document O is added to the vector database, \circledred{2} it is simultaneously passed as input to the LLM, and its KV cache is generated during the LLM prefill phase running on a GPU. \circledred{3} The computed KV cache, denoted as O's KV, is then stored in commodity flash storage for future retrieval. This step eliminates the need for recomputation, enabling efficient reuse of KV caches whenever the document is referenced again in future queries.

During inference, when a document is retrieved in response to a query, \MATKV retrieves the relevant precomputed KV caches instead of computing them on-the-fly, as shown in Figure~\ref{fig:matkv-archi}b. When a user submits a query Q, \circledblue{1} the inference system first performs a top-K search on the vector database to identify the most relevant documents. Instead of performing a computationally expensive prefill phase, \circledblue{2} the system retrieves the precomputed KV cache from flash storage. \circledblue{3} These retrieved KV caches are then loaded into GPU memory, where they are directly utilized for decoding, bypassing the need for KV recomputation for prefilling. This process effectively decouples the prefill phase from inference execution, enabling faster response times and higher throughput.

It is worth noting that \MATKV always places the precomputed KV caches of the retrieved documents before the user query in the prompt, as in typical RAG settings (Figure~\ref{fig:matkv-archi}b). 
This setup allows only the query to attend to the retrieved documents, and therefore the K and V matrices for the documents can be precomputed independently of any specific query.

A key question in adopting \MATKV is whether loading precomputed KV caches from storage instead of recomputing them on the GPU impacts LLM inference accuracy. Since MatKV treats each retrieved document independently, KV caches are precomputed separately, and documents always start at the first input position. This differs from conventional RAG inference, where multiple documents are concatenated in sequence, leading to varying positional embeddings and potential self-attention across documents. The absence of cross-document self-attention in \MATKV raises the question of whether it affects response quality.

To investigate this, we conducted preliminary experiments using 2WikiMultihopQA (LongBench)~\cite{2wikimqa, LongBench}, comparing vanilla inference, which computes KV caches in the conventional way on the GPU with full self-attention, to MatKV, which retrieves KV caches from SSD.

As shown in Table~\ref{tab:matkv-prelim}, \MATKV produces responses comparable to those obtained via vanilla inference (for more details, see Section~\ref{sec:accuracy}). One possible reason for \MATKV's improved accuracy is that vanilla inference processes multiple documents within a single context window, potentially leading to token loss, position bias, or diluted attention~\cite{lost_in_the_middle}. When documents are concatenated into a long context, tokens in the middle or near the cutoff may be dropped or receive less attention, thereby reducing accuracy. In contrast, MatKV loads each document’s KV cache independently, avoiding these issues and ensuring that all retrieved content remains equally accessible to the model. These results highlight a key insight: self-attention between the query and each retrieved document is sufficient for accurate generation, even in the absence of cross-document attention. This observation suggests that individual documents can independently contribute meaningful context to the final answer, challenging the necessity of performing attention across retrieved documents and offering new opportunities for simplifying retrieval-augmented architectures.

\subsection{Overlapping Prefill I/O with GPU Decoding}
In traditional architectures where both the prefill and decode phases rely entirely on GPU computation, these phases must be strictly serialized due to their shared dependence on GPU resources. To prevent interference between the two phases, it is common practice to schedule prefill and decode as separate tasks, even when processing multiple requests in batches~\cite{Sarathi}. However, this strict separation limits efficiency. The prefill phase—responsible for computing KV caches from retrieved documents—tends to saturate GPU resources due to the full forward pass through the model. In contrast, the decode phase, which generates tokens autoregressively, is generally less compute-intensive and utilizes the GPU far less aggressively. Without overlapping these phases, the GPU remains partially underutilized during decoding, and no useful work can be done while waiting for materialized KVs to be loaded from external storage.

In contrast, \MATKV eliminates the need for on-the-fly KV computation by replacing the prefill phase with storage reads from SSD to GPU memory, allowing for phase overlap. While the GPU is actively decoding the current batch, it can simultaneously prefetch KV pairs for the next batch from SSD, hiding the KV loading latency that would otherwise occur in a conventional prefill phase. This overlap-based execution minimizes idle GPU time and improves overall system throughput.

\begin{figure}[t]
  \includegraphics[width=\columnwidth]{figures/timeline.pdf}
  \caption{Overlapping in \MATKV. In conventional architectures, all phases are strictly serialized, whereas \MATKV enables concurrent KV loading and decoding.}
  \label{fig:matkv-overlap}
    \vspace{-3.4mm}
\end{figure}

Figure~\ref{fig:matkv-overlap} illustrates this difference, comparing the strictly serialized execution in conventional architectures with \MATKV's overlap-based approach, which disaggregates prefill and decode. In the traditional approach, the prefill phase which computes KV tensors for the given documents (D) and query (Q) (denoted as C(D,Q) in Figure~\ref{fig:matkv-overlap}) is strictly serialized with the decode phase. In contrast, \MATKV preloads KV pairs from flash storage while processing the previous batch, allowing overlap between KV loading and GPU computation. By enabling these phases to run concurrently, \MATKV reduces prefill-induced latency and enhances overall system efficiency (see Section~\ref{sec:batch} for more details).

\subsection{Advantages\label{sec:adv}}
\module{Latency and Throughput Improvements}
\MATKV offers multiple advantages, particularly in terms of inference speed, throughput, and power and cost efficiency. By replacing GPU compute time for processing input tokens with SSD read operations, \MATKV significantly shortens the prefill phase, leading to reduced overall inference latency. The extent of this reduction depends on the proportion of input tokens relative to the user query and output length. Additionally, \MATKV enables overlapping execution, where the GPU actively decodes the current batch while simultaneously prefetching KV pairs for the next batch from SSD. This effectively hides KV loading latency, preventing stalls and minimizing idle GPU time, and ensuring efficient resource utilization.

\module{Carbon-Efficient Prefill Phase}
\MATKV provides a substantial advantage in carbon-efficient inference by eliminating the need for GPU-intensive KV computation and replacing it with SSD-based KV retrieval. By offloading the prefill phase to SSDs, MatKV significantly reduces energy consumption. As mentioned in Section~\ref{sec:trading-com-for-sto}, for example, prefilling 1,204 tokens on an Nvidia H100 GPU consumes approximately 170 joules while Samsung 9100 Pro SSD requires just 0.14 joules, which is over 1,200× lower energy consumption. 

\subsection{{Discussions}}~\label{sec:discussion}

The primary goal of this paper is to establish, both analytically and empirically, that \MATKV is fundamentally superior to GPU recomputation in terms of cost and performance. To isolate and clearly demonstrate this core advantage, we deliberately adopted a simplified {\em Eager and Materialize-All} strategy, in which every ingested document is materialized on SSD regardless of its eventual access frequency. This design choice serves as a controlled baseline that removes the confounding effects of selective caching, eviction heuristics, or workload prediction. By doing so, we ensure that our evaluation focuses solely on the intrinsic trade-off between GPU recomputation and SSD-backed reuse of KV caches. Nevertheless, when considering practical deployment, additional challenges must be addressed, such as the cold start problem, selective materialization and eviction, and the storage overhead that influences the total cost of ownership (TCO).

\module {Cold Start Problem} Cold start is an inherent event because the first retrieval of a newly ingested document necessarily incurs the full prefill and materialization latency. While this can degrade user experience for that single request, our empirical workload analysis (Figure~\ref{fig::skewedaccess}) and the prior RAG study~\cite{RAGCache} demonstrate that queries are highly skewed toward a small fraction of hot documents. As a result, cold start events do not dominate overall performance in typical RAG scenarios. Even in high-ingestion environments such as news feeds or financial streams, the penalty can be bounded and mitigated by several mechanisms, including lazy materialization upon first retrieval, background precomputation using idle GPU cycles, or temporary DRAM buffering that serves initial requests while SSD materialization completes. These approaches ensure that cold start latency remains manageable rather than a fundamental limitation of \MATKV.

\module{Caching Policy} While the ten-day rule provides a principled break-even point showing that \MATKV is economically beneficial for documents with moderate access frequency, it is neither necessary nor efficient to persist all KV caches. Our analysis confirms that access distributions are highly skewed, with a small fraction of documents being retrieved repeatedly while most remain cold. In addition, as the query workload evolves over time, the set of repetitively accessed documents can change. This naturally motivates selective materialization policies that retain only hot documents, combined with eviction mechanisms triggered when SSD capacity is saturated. Simple recency- or frequency-based policies can be applied, while more advanced predictive methods can anticipate surging demand from workload trends such as emerging topics or breaking news. Such mechanisms transform \MATKV from a naïve store-everything prototype into a principled caching layer tailored to realistic retrieval workloads.

\module{Total Cost of Ownership (TCO)} The storage overhead of KV caches directly affects TCO. A naïve Materialize-All approach applied to billion-scale corpora and large models could consume terabytes or even petabytes of storage. However, three factors significantly mitigate this concern: first, selective caching reduces the total amount of materialized caches by orders of magnitude; second, compression techniques for KV caches~\cite{MiniCache,CacheGen, RocketKV,xKV} have demonstrated two- to four-fold reductions in size with negligible accuracy loss; and third, hierarchical storage architectures can tier KV caches across DRAM, SSD, and archival layers to balance cost and performance. Importantly, even under the conservative Materialize-All baseline, our ten-day rule analysis already shows that \MATKV is more economical than GPU recomputation. With selective caching, compression, and tiered storage, the effective cost can be reduced by at least an order of magnitude, further strengthening \MATKV's economic advantage.
\section{Implementation}\label{sec:impl}

\definecolor{green}{HTML}{008000} 

We have prototyped \MATKV by extending the Hugging Face Transformers library~\cite{HuggingFace-Transformer} with fewer than 500 code lines.

\module{Materializing KVs for RAG Objects}
Documents are first chunked into a predefined size (e.g., 1,024 tokens). Each chunk is then assigned a unique identifier and simultaneously stored in a vector database and on an SSD.

\begin{itemize}[leftmargin=10pt]
\item \texttt{get(O)}: Each chunk is stored in an in-memory vector database, ChromaDB~\cite{chromadb}. When storing chunks in ChromaDB, each chunk is transformed into an embedding vector using an embedding model (default: \texttt{all-MiniLM-L6-v2}~\cite{all-mini-lm}), which is customizable according to user requirements. Each embedding vector is stored along with its corresponding chunk identifier (\texttt{chunk\_id}), which enables the system to efficiently lookup and load the corresponding materialized KV tensors from the SSD later during inference.

\item \texttt{store\_KV(O's KV)}: Each chunk is processed through the LLM’s prefill phase through \texttt{get\_KV(O)} and \texttt{prefill\_KV(O)}. The resulting KV tensors are initially transferred from GPU memory to a CPU bounce buffer and then written asynchronously to the SSD as raw byte data using the \texttt{async\_io} operator provided by DeepNVMe, an optimized tensor I/O module of DeepSpeed~\cite{DeepSpeed}. Each precomputed KV tensor is stored as a file with the corresponding \texttt{chunk\_id} as its file name.

\item \texttt{delete(O)}: When an embedding for a document chunk (O) is removed from vector databases, its materialized KVCache also becomes stale and thus is also deleted from SSD using its \texttt{chunk\_id}.
\end{itemize}

\module{Avoiding Prefill Using KV in SSD}
\begin{itemize}[leftmargin=10pt]
    \item \texttt{load\_KV(O's KV)}. When retrieving relevant chunks, \MATKV uses the \texttt{chunk\_id} to asynchronously load the corresponding precomputed key-value tensors in the raw data bytes format from the SSD into CPU bounce buffer through DeepNVMe’s \texttt{async\_io} operator. The data are then copied from the CPU to GPU memory, allowing tensor computation with model.
    \item \texttt{put\_KV(O's KV)}. The loaded KV pairs are supplied to the model's \texttt{generate()} function (from the Hugging Face Transformers library) via the \texttt{past\_kv\_caches} parameter. By combining these precomputed KV pairs for the document with the query’s KV pairs, the system can skip the document’s prefill phase and proceed directly with autoregressive token generation.
\end{itemize}

\module{Overlapping Implementation}
We employ Python’s multiprocessing library to orchestrate two independent processes and effectively hide data-loading latency. The first process fetches query data, retrieves relevant chunks from vector database, prepares the batch’s KV cache of each chunk, and enqueues both into a queue shared between two processes. The second process retrieves these data from the queue, then executes the \emph{prefill} and \emph{decoding} steps for token generation. While the second process is generating tokens for the current batch, the loading process simultaneously prepares the KV cache for the subsequent batch. This parallel processing pipeline significantly reduces inference latency by overlapping I/O-bound and computation-bound operations.
\section{Evaluation\label{sec:eval}}
\subsection{Experimental Setup}\label{sec:eval-setup}

\module{Models and hardware settings} We primarily evaluate \MATKV on LLaMA~\cite{llama} 3.1 70B, as well as on LLaMA 3.2 3B and LLaMA 3.1 8B, to assess performance across a range of model sizes. All experiments are conducted using a H100 server configured with an NVIDIA H100 GPU with 80GB of HBM memory and a dual-socket Intel Xeon Gold 6442Y CPU with 2~TB of DRAM, and/or a single Nvidia RTX 4090 with 24GB of GPU memory. 

To ensure that the 70B model fits within the 80GB memory constraint of the H100, we use a 4-bit quantized version of LLaMA 3.1 70B, which reduces the model size from approximately 140GB (assuming float16) to around 35GB. This quantized version enables full model loading and inference on a single H100 without requiring model sharding or multi-GPU setups. The 3B and 8B models use float16 precision, as their parameter sizes are natively small enough to fit in both H100 and RTX 4090 configurations.

For KV storage, two types of SSDs are used. On the H100 GPU server, we use four Samsung 9100 Pro SSDs~\cite{9100Pro} (1TB each), configured in RAID-0 to provide high sequential read throughput of up to 58.8GB/s (14.7GB/s × 4). On the RTX 4090 setup, we use a single Samsung PM9A3 SSD~\cite{PM9A3} (1TB capacity, 6.5GB/s measured sequential read throughput). These SSDs store the materialized KV caches used by \MATKV in various retrieval and inference experiments.

\module{Datasets} Our evaluation utilizes two key datasets:
\begin{itemize}[leftmargin=10pt]
\item \emph{TurboRAG\cite{TurboRAG}.} We use dataset samples provided by TurboRAG to measure system latency and power consumption. In the dataset, the average number of tokens per query is 17.67, and the average number of tokens per document is 767.73. In experiments using a chunk size of 1,024, only chunks exceeding 1,024 tokens were considered to ensure experimental accuracy.

\item \emph{LongBench\cite{LongBench}.} To evaluate answer quality, we employ the \textsc{2WikiMultihopQA}~\cite{2wikimqa}, \textsc{HotpotQA}~\cite{hotpotqa}, and \textsc{TriviaQA}~\cite{triviaqa} datasets. The first two datasets require multi-hop reasoning, where answers must be derived by combining information from multiple documents, while the last one is evaluated in a few-shot setting.
\end{itemize}

\module{Performance Metrics} We measure three primary performance metrics:
\begin{itemize}[leftmargin=10pt]
    \item \emph{Latency.} We break down total response time into the following components:
    \begin{enumerate}
    \item \textbf{Load time}: The time required to load precomputed KV tensors from SSD into GPU memory.
    \item \textbf{Prefill time}: The duration from the completion of loading KV tensors until the generation of the first token.
    \item \textbf{Decode time}: The time spent generating tokens sequentially after the first token, until the sequence completion.
    \end{enumerate}
    \item \emph{Power Consumption.} We record power draw (in watts, W) during the experiment and integrate over time to obtain energy usage (in joules, J). 
    \item \emph{Answer Accuracy.} We quantify the correctness of generated responses using the F1 score.
\end{itemize}

\subsection{Basic Performance: Vanilla vs. \MATKV\label{sec:basic-perf}}

To evaluate the performance of \MATKV in comparison to the baseline GPU-only approach (denoted as Vanilla, i.e., full KV computation on GPUs), we conduct a series of experiments using the TurboRAG dataset~\cite{TurboRAG}. Each request consists of two retrieved chunks of 1,024 tokens each, along with a query of approximately 20 tokens. Based on this input, we generate a response with a maximum length of 20 tokens, which is typical in RAG workflow. We measure execution time across both the prefill and decode phases. In this subsection, we compare \MATKV against Vanilla across multiple performance dimensions, including sequential vs. batched execution, the impact of storage performance, the effect of overlap optimizations, and the power efficiency of \MATKV.

\begin{figure}[t]
  \includegraphics[width=0.9\columnwidth]{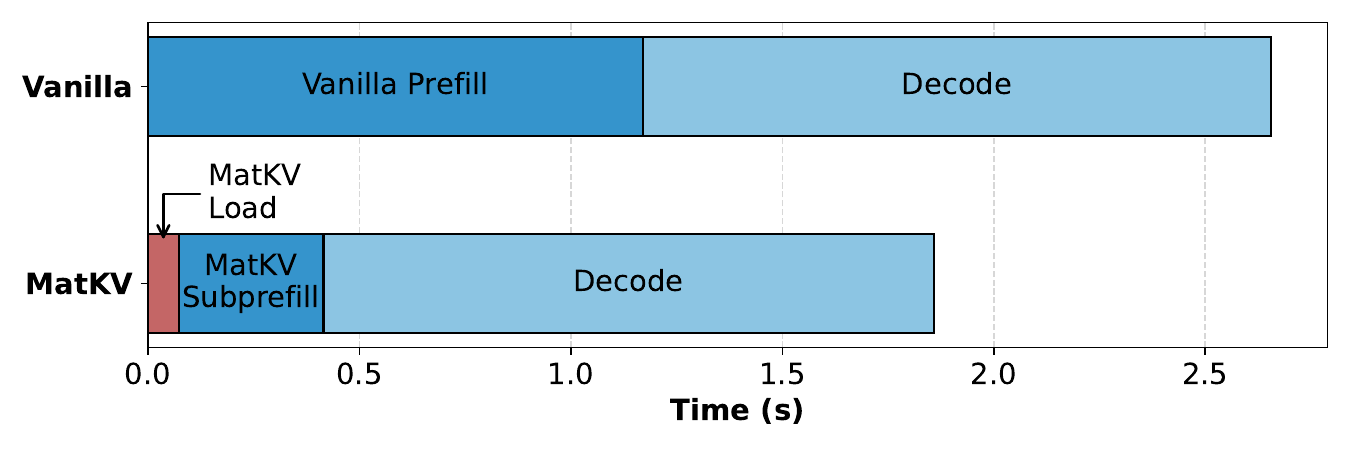}
  \centering
  \caption{Prefill and decoding latency comparison between \MATKV and Vanilla for a single request (LLaMA 3.1 70B). \MATKV reduces prefill time to less than half of Vanilla, demonstrating the efficiency of flash storage over GPU recomputation.}
  \label{fig:one-request}
\end{figure}

\subsubsection{Single-Request Performance Comparison} 
To assess the performance of \MATKV relative to Vanilla, we executed 1,024 requests sequentially, each consisting of 2,048 input tokens, 20 query tokens, and 20 answer tokens, measuring the total time spent in both the prefill and decode phases. The results are presented in Figure~\ref{fig:one-request}.

\module{Single-Request Performance of \MATKV} First of all, Figure~\ref{fig:one-request} demonstrates that \MATKV reduces the prefill time to less than half that of Vanilla. As intended, the high-speed performance of flash storage enables \MATKV to outperform full KV recomputation on a high-end H100 GPU, particularly for long input sequences. This result is highly promising, as one of the primary motivations for using high-end GPUs like the H100 in LLM inference is to accelerate the compute-intensive prefill phase. However, our findings show that \MATKV, which relies on inexpensive flash storage, not only replaces but even outperforms GPU-based prefill computation in terms of efficiency.
Nevertheless, since \MATKV still requires an additional prefill step for the query, as well as time for time-to-first-token and the subsequent answer token decoding, the relative performance improvement over Vanilla is approximately 70\%. As will be demonstrated later, by executing multiple requests in batches rather than sequentially, we can amortize the latency contribution of the decode phase, further emphasizing the benefits of \MATKV in optimizing prefill efficiency.

\begin{table}[t]
    \centering
    \caption{Impact of Storage Performance (128 Requests)} 
    \label{tab:storage}
    \begin{tabular}{c|cc}
    \toprule
    Storage  & \makecell{Per-Request Average} & \makecell{Total Load} \\
            & \makecell{Load Time (s)} & \makecell{Time (s)} \\
    \midrule 
    One 9100 Pro SSD & 0.093 & 11.97 \\
    Four RAIDed SSDs & 0.027 & 3.53 \\ 
    DRAM & 0.006 & 0.77 \\
    \bottomrule
    \end{tabular}
      \vspace{-3mm}
\end{table}

\module{Impact of Storage Performance on \MATKV} The flash storage used in the above experiment consists of four Samsung 9100 Pro SSDs, configured in software RAID 0, providing a sequential read performance of 30 GB/s. To evaluate the impact of storage performance on \MATKV, we conducted the same experiment under two additional configurations: a single 9100 Pro SSD and DRAM storage. In the DRAM storage configuration, KVs were preloaded into DRAM, and only the loading time to HBM was measured using the {\tt aio} method. The results are presented in Table~\ref{tab:storage}. As expected, the loading time in \MATKV is highly sensitive to storage performance, exhibiting significant variations depending on the underlying storage. Although loading from DRAM is faster than from SSDs, the high cost of DRAM makes it impractical for storing KVs in real-world applications as discussed in Section~\ref{sec:trading-com-for-sto}.

\subsubsection{Impact of Batched Execution on \MATKV Performance\label{sec:batch}}

\begin{figure}[t]
  \includegraphics[width=0.77\columnwidth]{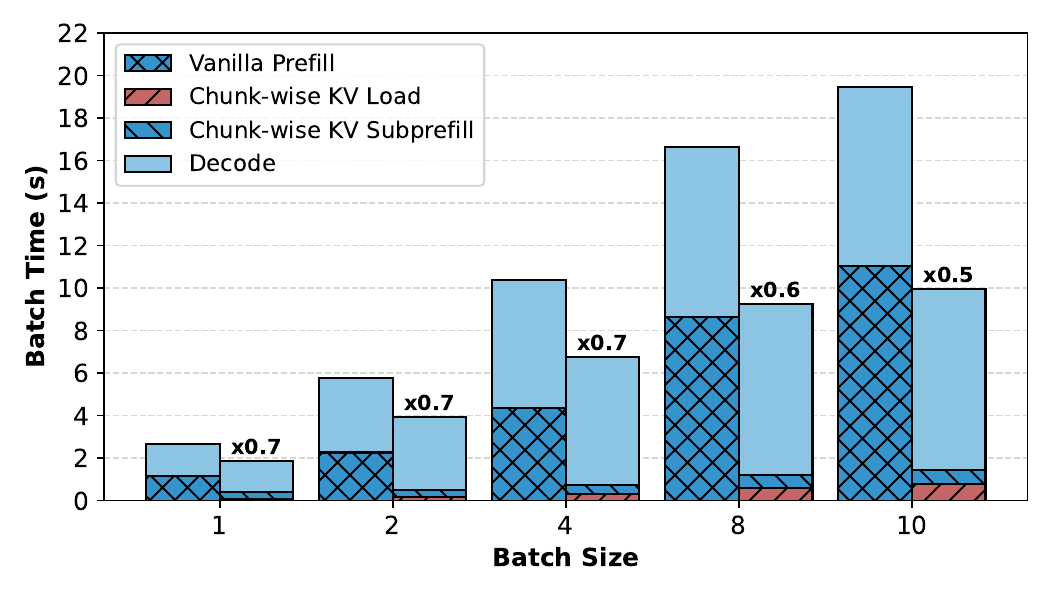}
  \centering
  \caption{Batch inference performance of Vanilla vs. \MATKV for 200 requests with batch sizes from 1 to 10 in LLaMA 3.1 70B. \MATKV reduces prefill time significantly, achieving up to 2× speedup as batch size increases.}
  \label{fig:batch-perf}
    \vspace{-3.4mm}
\end{figure}

In the above, we have demonstrated that \MATKV outperforms full KV compute on an H100 GPU during the prefill phase. However, since decode time remains the dominant factor in overall LLM inference latency, the advantages of \MATKV were not as pronounced. The decoding phase is autoregressive and memory-access intensive rather than compute-intensive, making single-request execution significantly slower compared to batched processing~\cite{PagedAtt}. Fortunately, LLM serving systems including Hugging Face Transformers and vLLM~\cite{{PagedAtt,HuggingFace-Transformer}}  support the batched inference: by executing multiple requests concurrently in a batch, GPU core parallelism can be effectively leveraged to improve decoding throughput~\cite{PagedAtt, Sarathi}.

To evaluate how \MATKV performs under batched execution, we process 200 requests (each with 2,048 input tokens, 20 query tokens, and 20 answer tokens) on both Vanilla and \MATKV, measuring prefill and decode time. Furthermore, to analyze the impact of varying the batch size, we repeat the experiment while increasing the number of requests per batch from 1 to 10. The results are presented in Figure~\ref{fig:batch-perf}.
As the number of requests per batch increases, the prefill time scales linearly, whereas the decode time grows sublinearly due to improved parallelism. Consequently, when the batch size exceeds 8 requests per batch, the decode phase no longer dominates the total execution time, and the prefill phase becomes the primary factor to inference latency. This means that \MATKV can significantly reduce computation time as batch size increases.

\begin{figure}[t]
    \centering
    \begin{subfigure}{0.49\linewidth}
        \centering
        \includegraphics[width=0.9\linewidth]{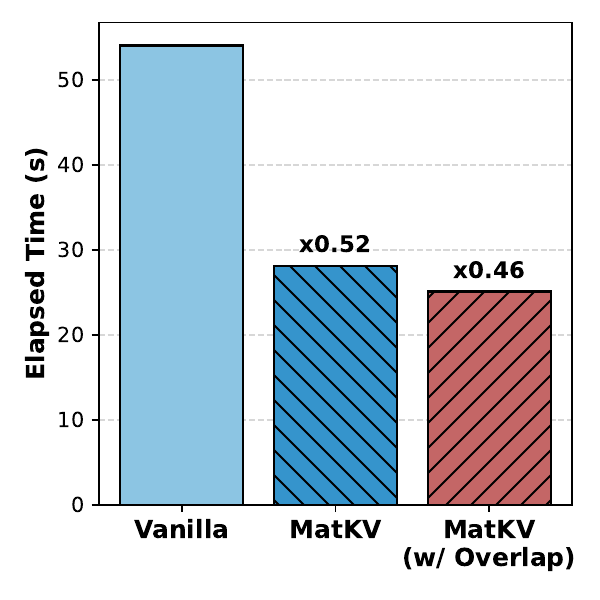}
        \caption{LLaMA 3.1 8B (batch of 32)}
        \label{fig:overlap-8b}
    \end{subfigure}
    \hfill
    \begin{subfigure}{0.49\linewidth}
        \centering
        \includegraphics[width=0.9\linewidth]{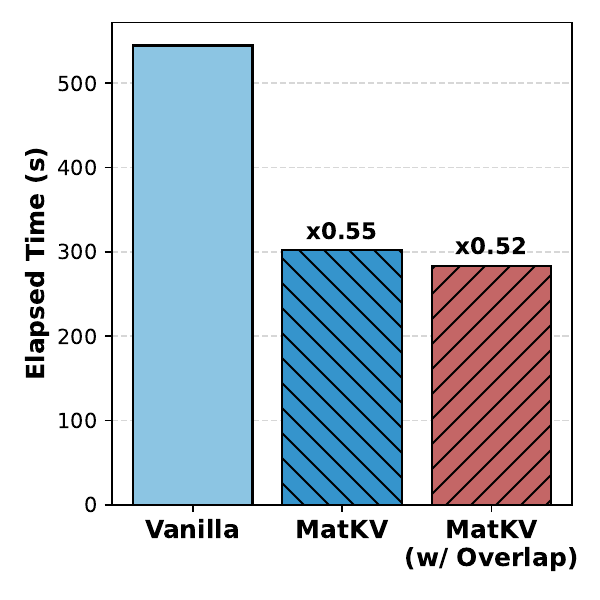}
        \caption{LLaMA 3.1 70B (batch of 8)}
        \label{fig:overlap-70b}
    \end{subfigure}
    \caption{Impact of overlapped prefill and decode execution on inference time for LLaMA 3.1 8B and 70B, respectively. \MATKV with Overlap achieves up to a 2× speedup over Vanilla by overlapping decoding with SSD-based KV loading.}
    \label{fig:overlap}
\end{figure}

As discussed in Section~\ref{sec:design}, \MATKV enables the overlap of the decode phase of the current batch with the prefill phase of the next batch. To quantitatively evaluate the impact of this overlapping execution — specifically, the concurrent execution of GPU decode computation and SSD I/O for KV loading — we conduct experiments using Vanilla, basic \MATKV, and overlapped \MATKV versions. Additionally, to assess the effect across different LLM models, we repeat the same experiment on two LLaMA 3.1 models: 8B and 70B. The results are presented in Figure~\ref{fig:overlap}.

Figure~\ref{fig:overlap} clearly confirms that, for both models, overlapped \MATKV consistently achieves a 2$\times$ speedup over Vanilla. Note that the y-axis value in Figure~\ref{fig:overlap-70b} is 10× larger than that in Figure~\ref{fig:overlap-8b}. This reflects that the larger 70B model need significantly higher inference cost. Compared to basic \MATKV, the relative improvement of the overlapped version is less pronounced. This is because overlapping primarily hides the KV loading time from SSD, whereas the decode phase remains the dominant execution bottleneck rather than the prefill phase. However, as we will demonstrate later, when the input size increases — leading to a higher proportion of time spent in prefill — the performance improvement of overlapped \MATKV over basic \MATKV will become more significant.

\subsubsection{Energy Consumption}

\MATKV replaces power-intensive GPU compute in the prefill stage with power-efficient flash I/O, thus reducing the power-consumption.  In this section, we demonstrate empirically how much energy can be saved by \MATKV for RAG workflows. 

\begin{table}[t]
    \caption{System-wide Power Consumption}
    \label{tab:power}
    \centering
    \begin{tabular}{c|cc|c|c}
    \toprule
    System  & \makecell{Peak} & \makecell{Average} & \makecell{ Time} & \makecell{Total} \\
    Configuration  & \makecell{(W)} & \makecell{(W)} & \makecell{Taken (s)} & \makecell{Power (kJ)} \\
    \midrule 
    Vanilla & 1,256 & 1,038 & 546 & 566 \\
    MatKV & 1,124 & 947 & 306 & 289 \\ 
    MatKV (w/ Overlap) & 1,241 & 979 & 285 & 279 \\
    \bottomrule
    \end{tabular}
\begin{tablenotes}
\small
\item * The idle system power was measured as about 550W
\end{tablenotes}
  \vspace{-3mm}
\end{table}

\begin{figure*}[t]
    \centering
    \begin{subfigure}[b]{0.48\textwidth}
        \centering
        \includegraphics[width=0.9\columnwidth]{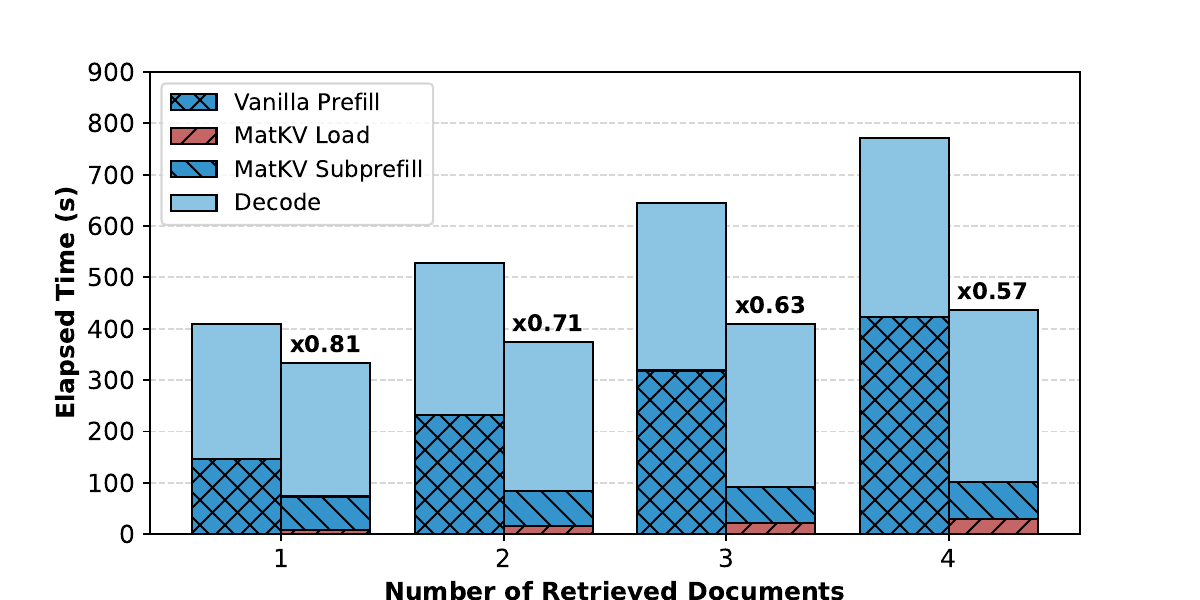}
        \caption{Varying Input Size}
        \label{fig:input-size}
    \end{subfigure}
    \hfill
    \begin{subfigure}[b]{0.48\textwidth}
        \centering
        \includegraphics[width=0.9\columnwidth]{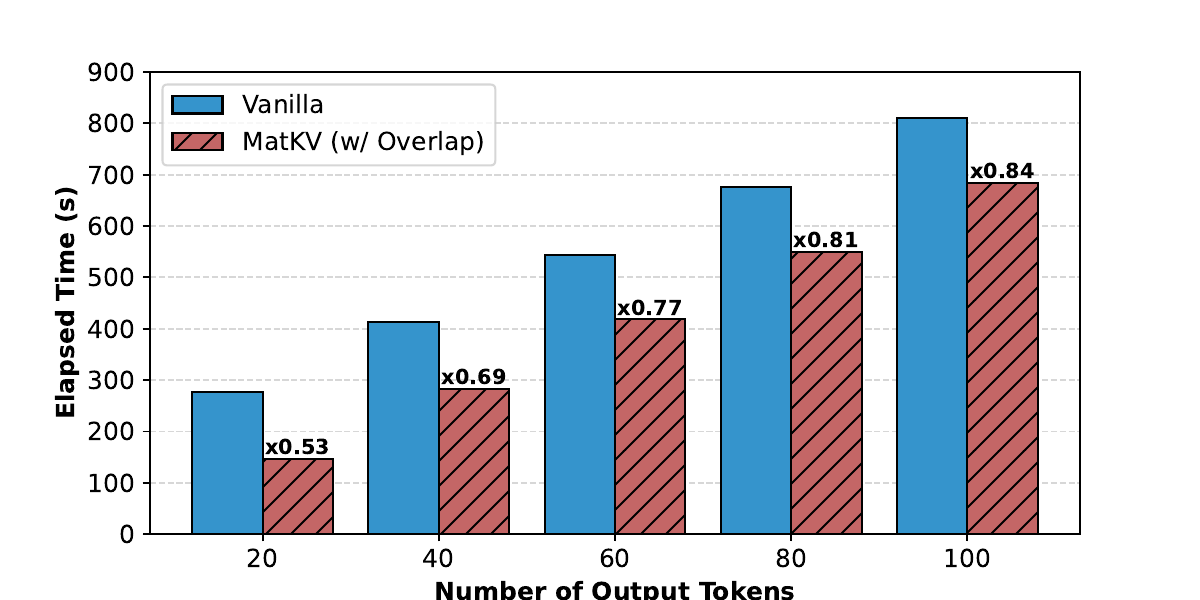}
        \caption{Varying Output Size}
        \label{fig:output-size}
    \end{subfigure}
    \caption{Impact of input and output size on inference time. (a) Increasing the number of retrieved documents (input size) improves MatKV's relative speedup over Vanilla by accelerating loading and subprefill phases. (b) Increasing output length reduces MatKV's relative speedup as decode time dominates, but \MATKV remains consistently faster than Vanilla.}
    \label{fig:input-output-size}
      \vspace{-3mm}
\end{figure*}

\module{System-wide Power Consumption} To measure the total system power consumption, including all components of the H100 GPU server, we utilize IPMI (Intelligent Platform Management Interface)~\cite{IPMI}, a widely used protocol for remote system management. IPMI operates independently of the system's CPU and operating system, making it a reliable tool for measuring overall server power consumption. 
At idle, the H100 server consumes 550W, with the CPU and DRAM each drawing approximately 90W, while four RAID-configured SSDs collectively consume around 30W.

To compare the power consumption of Vanilla (full KV computation) and \MATKV under actual workloads, we execute the previously used RAG workflow (256 batches, batch size = 8) on the H100 GPU server. During execution, we measure peak and average power consumption (in watts) and record the total execution time required to complete 256 requests, along with the corresponding total energy consumption. The results, summarized in Table~\ref{tab:power}, show that both Vanilla and overlapped \MATKV exhibit similar peak power consumption (second column), while basic \MATKV shows slightly lower peak usage. In terms of average power consumption (third column), both \MATKV variants consume approximately 10\% less power than Vanilla, primarily due to their ability to bypass the compute-intensive prefill phase. More significantly, since overlapped \MATKV completes execution twice as fast as Vanilla (fourth column), its total energy consumption is less than half that of Vanilla (fifth column), demonstrating substantial energy efficiency.

\module{Power Consumption by GPU} To precisely measure the power consumption attributed solely to the GPU, rather than the entire system, we monitor GPU power usage during the same experiment. Using Nvidia’s SMI (System Management Interface)~\cite{nvidia-smi}, we measure the peak and average power consumption of the H100 GPU while processing 256 requests. Additionally, we compute the total energy consumption required to complete the workload. The results are summarized in Table~\ref{tab:power-gpu}.  
Across all three configurations, the peak power consumption (second column) reaches the GPU power cap of 350W, while the average power consumption (third column) remains similar across configurations. Since the overlapped \MATKV version completes execution twice as fast as the full KV compute (Vanilla) approach, its total power consumption is approximately half that of Vanilla.

\begin{table}[t]
    \caption{GPU Power Consumption} 
    \label{tab:power-gpu}
    \centering
    \begin{tabular}{c|cc|c|c}
    \toprule
    System  & \makecell{Peak} & \makecell{Average} & \makecell{ Time} & \makecell{Total} \\
    Configuration  & \makecell{(W)} & \makecell{(W)} & \makecell{Taken (s)} & \makecell{Power (kJ)} \\
    \midrule 
    Vanilla & 353 & 340 & 546 & 185 \\
    MatKV  & 355 & 322 & 306 & 98 \\ 
    MatKV (w/ Overlap) & 356 & 336 & 285 & 95 \\
    \bottomrule
    \end{tabular}
\begin{tablenotes}
\small
\item * The idle GPU power was measured as about 50W
\end{tablenotes}
  \vspace{-3mm}
\end{table}

\subsection{Other Performance Comparison} 

\subsubsection{Varying Input and Output Length}  
To further analyze how Vanilla and \MATKV perform under different input and output sizes, we conduct the same experiment while varying the input and output size.

\module{Varying Input Length} First, to verify the effect of different input size, we run the experiment while varying the number of input chunks from 1 to 4. The batch size is set to 1 for both systems, and the non-overlapped version of \MATKV was used. The results are presented in Figure~\ref{fig:input-size}. 

In the case of Vanilla, the prefill time increases proportionally to the input size, but the decode time grows sublinearly due to batch parallelism.
In contrast, \MATKV significantly accelerates the loading and subprefill phases, completing them much faster than the prefill phase of Vanilla. Furthermore, as the batch size increases, the proportion of the decode phase relative to the prefill phase in Vanilla decreases. Consequently, the relative performance gap between \MATKV and Vanilla (annotated on the bars in the graph) is, as intended, widened with large input size.

\module{Varying Output Length} Conversely, to examine how \MATKV works with longer answer but fixed input size, we conducted the same experiment while varying the generated output length from 20 to 100 tokens. The results are presented in Figure~\ref{fig:output-size}. Since the decode phase obviously takes longer with larger answer, the decode time will accordingly dominate the inference time, leading to a reduced relative performance gain of \MATKV. However, despite the diminishing relative improvement, \MATKV still consistently outperforms Vanilla, demonstrating its efficiency even under longer output sizes.

\subsubsection{Effect of \MATKV By Model Size} 

\begin{figure*}[t]
    \centering
    \begin{subfigure}[b]{0.48\textwidth}
        \centering
        \includegraphics[width=0.8\columnwidth]{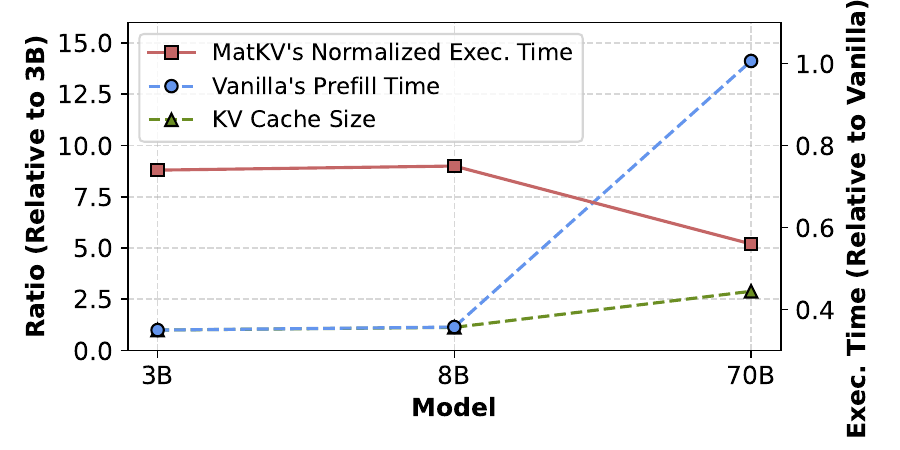}
        \caption{Input Token Number = 1,024}
        \label{fig:model-size-a}
    \end{subfigure}
    \hfill
    \begin{subfigure}[b]{0.48\textwidth}
        \centering
        \includegraphics[width=0.8\columnwidth]{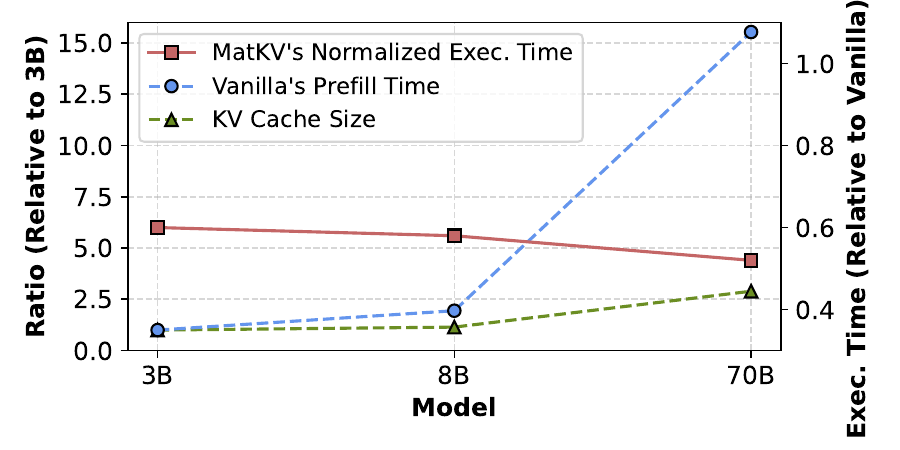}
        \caption{Input Token Number = 2,048}
        \label{fig:model-size-b}
    \end{subfigure}
    \caption{Impact of model size on \MATKV’s efficiency. As model size increases, the compute cost for KV computation (blue) grows faster than KV cache size (green), amplifying \MATKV’s benefit (red). This trend remains consistent across different input token lengths, confirming \MATKV’s scalability.}
    \label{fig:model-size}
      \vspace{-3.4mm}
\end{figure*}

The prefill time and KV cache size vary across different LLMs due to differences in model architecture, parameter count, and memory footprint. Consequently, the relative benefit of \MATKV over full KV computation is model-dependent. To evaluate the effectiveness of \MATKV across different LLMs, we measure the average prefill time and KV size per RAG object while processing an identical workload consisting of 256 requests. We conduct experiments using three LLaMA models, including 3B, 8B, and 70B, to assess how model size influences KV storage requirements and compute overhead. Additionally, we compute the relative performance gain of \MATKV over full KV computation across these models. The results are presented in Figure~\ref{fig:model-size-a}.

The result from Figure~\ref{fig:model-size-a} clearly shows that as the model size grows, the compute required for calculating KVs (the {\textcolor{blue}{blue}} dotted line) amplifies more steeply than the KV size (the 
{\textcolor{green}{green}}  dotted line) does. Note that both lines are plotted based on the left Y-axis values. As a result, the benefit of \MATKV which can hide the prefill latency will be bigger (the {\textcolor{red}{red}} line which is plotted based on the right Y-axis values). 

To further investigate whether the benefits of \MATKV persist across varying input token sizes, we repeat the same experiment while increasing the number of input tokens from 1,024 to 2,048.  The results are presented in Figure~\ref{fig:model-size-b}. As expected, both prefill time and KV size increase proportionally with the input token length. However, the relative benefit of \MATKV remains consistent, demonstrating that \MATKV’s efficiency gain is independent of both model size and input token length. This confirms that \MATKV effectively scales across different LLM configurations and input sizes.

\subsubsection{Prefill with \MATKV and Decode with low-end GPU\label{sec:low-end-gpu}}
\begin{figure}[t]
  \centering
  \includegraphics[width=0.9\columnwidth]{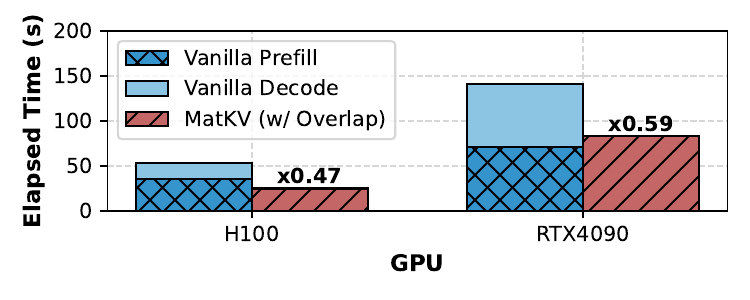}
  \caption{\MATKV vs. full KV recompute on H100 and RTX 4090. Despite being 30× cheaper, \MATKV on RTX 4090 is only 1.5× slower than full KV computation on H100, demonstrating cost-efficient inference on low-end GPUs.}
  \label{fig:matkv-across-GPUs}
    \vspace{-3.4mm}
\end{figure}
Since \MATKV decouples the prefill and decode phases, it allows the decode phase to be executed on different GPUs. Given that decoding is not compute-intensive, the decoding time on a low-end GPU does not take much longer than a high-end GPU~\cite{Sarathi}. This opens up a cost-effective deployment strategy where a high-end GPU computes and stores KV caches on SSD, while a low-end GPU handles the decoding phase, significantly improving cost-performance efficiency.

To validate this approach, we repeat the same experiment used for Figure~\ref{fig:batch-perf}, which uses 200 requests with an input token size of 1,024, and compare the performance of full KV computation and \MATKV across two GPU servers: the high-end Nvidia H100 and the low-end RTX 4090. As of March 2025, the H100 server, priced at approximately \$50K, is 30× more expensive than the RTX 4090, which costs around \$1.6K. Due to the differences in parallelism between the two GPUs, the 256 requests are processed with a batch size of 32 on the H100 and a batch size of 2 on the RTX 4090, respectively. The results are presented in Figure~\ref{fig:matkv-across-GPUs}.

Figure~\ref{fig:matkv-across-GPUs} shows that, compared to the full KV recompute on H100, \MATKV on RTX 4090 is only 1.5$\times$ slower while Vanilla is 3$\times$ slower. This demonstrates the feasibility of offloading the decoding phase to low-end GPUs without significant performance degradation, reinforcing the cost-efficiency of \MATKV. Given that practitioners try even CPU-based inference for RAG applications due to the cost and power concerns~\cite{CPUinferencing}, \MATKV makes CPU-only inference practical.

\begin{table}[t]
    \caption{\MATKV Accuracy (F1)} 
    \label{tab:accuracy}
    \centering
    \begin{tabular}{l|r|r|r}
    \toprule
    Dataset & Vanilla & MatKV & CacheBlend \\ 
    \midrule 
    2WikiMQA & 0.278 & 0.241 & 0.253 \\
    TriviaQA & 0.786 & 0.812 & 0.826 \\
    HotpotQA & 0.353 & 0.302 & 0.309 \\
    \bottomrule
    \end{tabular}
    \vspace{-3mm}
\end{table}

\subsubsection{\MATKV Accuracy\label{sec:accuracy}}

To evaluate the impact of these factors on RAG performance, we measure QA accuracy on three datasets from the LongBench benchmark~\cite{LongBench}: 2WikiMultihopQA~\cite{2wikimqa}, TriviaQA~\cite{hotpotqa}, and HotpotQA~\cite{triviaqa}. Each dataset consists of 200 queries, and the documents are chunked into segments to fit the RAG setting. The evaluation is conducted based on responses generated using the top-5 retrieved documents. In addition, we compare against CacheBlend~\cite{CacheBlend}, which performs recomputation on 18\% of the retrieved KV cache.
An additional experiment evaluated \MATKV on chat history. On the Locomo dataset~\cite{locomo}, long histories were chunked and processed via RAG. Accuracy was 0.423 for Vanilla and 0.419 for \MATKV, showing negligible degradation.

Beside accuracy, we also compare \MATKV with CacheBlend in terms of speed. \MATKV outperforms CacheBlend by far mainly due to its optimization in loading materialized KV Caches from SSD.
Concretely, the loading time from SSD in \MATKV is 37\% faster that that in CacheBlend and the time taken to first token (TFFT) is 41\% faster. TFFT in CacheBlend is relatively longer than loading time because of its blending (i.e., additional cross-attentioning time).
\section{Related Work}~\label{sec:related} 

In that \MATKV replaces the GPU compute with flash storage I/O operation and reduce the prefill time with fast flash storage, several types of existing works are related: scalable KV computation~\cite{PagedAtt, FlashAtt}, KV cache size reduction~\cite{MQA,GQA}, DRAM-based KV caching~\cite{RAGCache,TurboRAG,PromptCache,CAG}, and overlapping KV cache transfer with GPU computation~\cite{dejavu, InfiniGen} Also, in that it materialize the precomputed result in storage to avoid recomputation, \MATKV is close to the materialized views in database community~\cite{MV,MV2}.

\module{Scalable KV Computation} 
Numerous approaches~\cite{PagedAtt,FlashAtt} have been developed to improve scalability and memory utilization, thereby accelerating KV computation. FlashAttention~\cite{FlashAtt} enhances self-attention efficiency by reducing memory overhead and optimizing computational speed. PagedAttention~\cite{PagedAtt} introduces a memory-efficient mechanism designed to handle large-scale context windows by dynamically managing KV cache across GPU and CPU memory. However, the prefill phase remains a major bottleneck in LLM services, particularly in RAG workloads where long input sequences are repeatedly processed~\cite{CacheBlend}. \MATKV eliminates the need for redundant prefill by storing and retrieving precomputed KVs using fast but inexpensive flash storage, leading to superior throughput and lower power consumption compared to GPU servers, as demonstrated in Section~\ref{sec:eval}. 

\module{KV Cache Size Reduction} To mitigate the growing memory footprint of KV caches in LLM inferences, various techniques have been proposed to reduce KV size.  Multi-Query Attention~\cite{MQA} and Group-Query Attention~\cite{GQA} reduce KV size by sharing key-value pairs across multiple queries, while compression-based approaches such as MiniCache, CacheGen, RocketKV, and xKV~\cite{MiniCache,CacheGen,RocketKV, xKV} further reduce storage overhead by storing KV pairs in a more compact format. However, these techniques do not reduce KV compute time and, in some cases, may even introduce overhead due to additional processing required. Unlike existing KV size reduction schemes, \MATKV benefits from both reduced memory usage and faster loading time, thereby lowering overall inference latency.

\module{KV Caching} 
Several KV caching schemes have been proposed to avoid redundant prefill computation by storing and retrieving precomputed KV caches from DRAM memory. CAG, and PromptCache~\cite{CAG,PromptCache} precompute frequently accessed prompts, store them in DRAM, and retrieve them when necessary. 
Additionally, KV caching techniques specifically designed for RAG workloads, such as RAGCache and TurboRAG~\cite{RAGCache,TurboRAG}  optimize KV reuse to enhance efficiency in dynamic retrieval scenarios. While these approaches reuse KV caches instead of recomputing them, making the concept behind \MATKV not entirely novel, they still rely on DRAM-based caching, which presents scalability and cost limitations.

In contrast, \MATKV's key contribution lies in demonstrating that materializing KV caches on commodity flash storage is not only more economical but also faster than performing prefill computation on enterprise-grade GPU servers. By shifting KV storage to high-performance flash rather than DRAM as discussed in Section~\ref{sec:raginfandcomp}, \MATKV enables a scalable, cost-effective, and high-throughput KV retrieval strategy. In that SSDs are also used as as storage medium, CacheBlend~\cite{CacheBlend} is closest to \MATKV. Furthermore, CacheBlend suggests how to improve the accuracy by blending (that is, cross-attention) multiple KV Caches which are independently prefilled and thus outperforms \MATKV slightly in terms of accuracy. However, as demonstrated in Section~\ref{sec:eval}, \MATKV significantly outperforms CacheBlend in terms of speed, due to its optimization on loading KV Caches from SSDs. Additionally, we suggest that with the overlaped execution of prefill and decode phase, SSD speed is enough to hide the loading latency.

\module{Overlapping KV cache loading with GPU computation} 
Prior works such as DejaVu~\cite{dejavu} and InfiniGen~\cite{InfiniGen} load offloaded KV caches during decoding, overlapping the loading process with GPU computation. Notably, InfiniGen overlaps only with the feed-forward network (FFN) layer due to the inherent nature of its approach.
In contrast, \MATKV overlaps the loading of precomputed KV caches for RAG documents during the prefill phase, rather than during decoding.

\module{Materialized Views}  The idea behind \MATKV, which trades compute for storage, is inspired by materialized views from the database community~\cite{MV,MV2}. Materialized views precompute query results and store them, allowing subsequent queries to be answered directly from the stored view rather than recomputing results from base tables. This approach significantly reduces computation overhead, aligning with \MATKV’s strategy of precomputing and storing KV caches to eliminate redundant prefill computation.
\section{Conclusion and Future Work}
To accelerate LLM inference and reduce costs, particularly in RAG environments, we propose \MATKV. It suggests to trade the GPU compute for the flash storage in prefilling input tokens in RAG workflow. We have demonstrated that \MATKV achieves better performance and carbon efficiency than GPU-only RAG inference when executed using state-of-the-art GPUs and flash memory SSDs. Additionally, by offloading materialized KVs to low-end GPUs, \MATKV enables a cost-effective alternative for LLM inference.

Two practical directions remain. First, while our evaluation results in Section~\ref{sec:eval} suggest that simply concatenating independently computed KVs does not trade accuracy severely, further study is needed to identify which tasks this holds for and where it breaks down. We also plan to extend KV reuse to other generative tasks, such as agentic AI~\cite{AgenticAI-Inf-Size}, text-to-SQL~\cite{Text2SQL-prompt-cache}, and in-context learning, which often involve repeated use of long texts (e.g., tool docs, demonstrations).

\section*{Acknowledgment}
This work was supported in part by the National Research Foundation of Korea (NRF) grant (MSIT) (RS-2024-00414981, RS-2025-00560762); by the Health and Medical R\&D Program of the Ministry of Health and Welfare (RS-2025-25455059); by the Institute of Information \& Communications Technology Planning \& Evaluation (IITP) grant (MSIT) (RS-2024-00338140); by Samsung Electronics; and by research facilities provided by the Samsung Memory Research Center (SMRC) and the M-square Lab.

\newpage
\section*{AI-Generated Content Acknowledgement}
The authors declare that no generative AI tools were used to generate any text, figures, images, or code in this paper. All content was created entirely by the authors.

\bibliographystyle{IEEEtran}
\bibliography{bibliography}

@misc{CAG,
      title={Don't Do RAG: When Cache-Augmented Generation is All You Need for Knowledge Tasks}, 
      author={Brian J Chan and Chao-Ting Chen and Jui-Hung Cheng and Hen-Hsen Huang},
      year={2024},
      eprint={2412.15605},
      archivePrefix={arXiv},
      primaryClass={cs.CL},
      url={https://arxiv.org/abs/2412.15605}, 
}

@inproceedings{CacheBlend,
author={Jiayi Yao and Hanchen Li and Yuhan Liu and Siddhant Ray and Yihua Cheng and Qizheng Zhang and Kuntai Du and Shan Lu and Junchen Jiang},
title = {CacheBlend: Fast Large Language Model Serving for
RAG with Cached Knowledge Fusion},
year = {2025},
booktitle = {Proceedings of the Twentieth European Conference on Computer Systems (EuroSys '25)}
}

@inproceedings{CacheGen,
author = {Liu, Yuhan and Li, Hanchen and Cheng, Yihua and Ray, Siddhant and Huang, Yuyang and Zhang, Qizheng and Du, Kuntai and Yao, Jiayi and Lu, Shan and Ananthanarayanan, Ganesh and Maire, Michael and Hoffmann, Henry and Holtzman, Ari and Jiang, Junchen},
title = {CacheGen: KV Cache Compression and Streaming for Fast Large Language Model Serving},
year = {2024},
booktitle = {Proceedings of the ACM SIGCOMM 2024 Conference},
pages = {38–56},
series = {ACM SIGCOMM '24}
}

@misc{CPUinferencing,
    author = {Kian Mohadjerin and Cédric Bulteel},
    title     = {CPU Inferencing: Why not?}, 
    howpublished = {\url{https://www.nscale.com/blog/cpu-inferencing-why-not?utm_source=chatgpt.com}},
    Month = {October},
    year = {2024}
}

@misc{DeepSeek,
    author = {DeepSeek},
    title     = {DeepSeek API introduces Context Caching on Disk, cutting prices by an order of magnitude}, 
    howpublished = {\url{https://api-docs.deepseek.com/news/news0802}},
    Month = {August},
    year = {2024}
}

@article{fiveminrule,
author = {Gray, Jim and Putzolu, Franco},
title = {The 5 minute rule for trading memory for disc accesses and the 10 byte rule for trading memory for CPU time},
year = {1987},
issue_date = {Dec. 1987},
publisher = {Association for Computing Machinery},
address = {New York, NY, USA},
volume = {16},
number = {3},
issn = {0163-5808},
journal = {SIGMOD Record},
month = dec,
pages = {395–398},
numpages = {4}
}

@misc{MQA,
      title={Fast Transformer Decoding: One Write-Head is All You Need}, 
      author={Noam Shazeer},
      year={2019},
      eprint={1911.02150},
      archivePrefix={arXiv},
      primaryClass={cs.NE},
      url={https://arxiv.org/abs/1911.02150}, 
}

@inproceedings{GQA,
title={{GQA}: Training Generalized Multi-Query Transformer Models from Multi-Head Checkpoints},
author={Joshua Ainslie and James Lee-Thorp and Michiel de Jong and Yury Zemlyanskiy and Federico Lebron and Sumit Sanghai},
booktitle={The 2023 Conference on Empirical Methods in Natural Language Processing},
year={2023},
url={https://openreview.net/forum?id=hmOwOZWzYE}
}

@misc{HuggingFace-Transformer,
    author = {HuggingFace},
    title     = {HuggingFace Transformer}, 
    howpublished = {\url{https://github.com/huggingface/transformers}},
    Month = {January},
    year = {2025}
}

@misc{all-mini-lm,
    author = {Sentence Transformers},
    title     = {sentence-transformers/all-MiniLM-L6-v2}, 
    howpublished = {\url{https://huggingface.co/sentence-transformers/all-MiniLM-L6-v2}},
    Month = {January},
    year = {2025}
}

@misc{DeepSpeed,
    author = {DeepSpeed},
    title = {Using DeepNVMe for simple file reads and writes involving CPU/GPU tensors},
    howpublished = {\url{https://github.com/deepspeedai/DeepSpeedExamples/tree/master/deepnvme/file_access}},
    Month = {February},
    year = {2025}
    }

@article{LongBench,
  title={Longbench: A bilingual, multitask benchmark for long context understanding},
  author={Bai, Yushi and Lv, Xin and Zhang, Jiajie and Lyu, Hongchang and Tang, Jiankai and Huang, Zhidian and Du, Zhengxiao and Liu, Xiao and Zeng, Aohan and Hou, Lei and others},
  journal={arXiv preprint arXiv:2308.14508},
  year={2023}
}

@misc{InfServing,
      title={LLM Inference Serving: Survey of Recent Advances and Opportunities}, 
      author={Baolin Li and Yankai Jiang and Vijay Gadepally and Devesh Tiwari},
      year={2024},
      eprint={2407.12391},
      archivePrefix={arXiv},
      primaryClass={cs.DC},
      url={https://arxiv.org/abs/2407.12391} 
}

@misc{IPMI,
    author = {Wikipedia},
    title     = {Intelligent Platform Management Interface}, 
    howpublished = {\url{https://en.wikipedia.org/wiki/Intelligent_Platform_Management_Interface}},
    Month = {January},
    year = {2025}
}

@misc{nvidia-smi,
    author = {NVIDIA},
    title = {System Management Interface},
    howpublished =
    {\url{https://developer.nvidia.com/system-management-interface}},
    Month = {January},
    year = {2025}
}

@misc{llamaindex-chunksize,
    author = {LlamaIndex},
    title = {Basic Strategies},
    howpublished ={\url{https://docs.llamaindex.ai/en/stable/optimizing/basic_strategies/basic_strategies}},
    Month = {January},
    year = {2025}
}

@misc{openai-chunksize,
author = {OpenAI},
title = {Assistants File Search},
howpublished =
{\url{https://platform.openai.com/docs/assistants/tools/file-search}},
Month = {March},
year = {2025}
}

@article{llama,
  title={The llama 3 herd of models},
  author={Grattafiori, Aaron and Dubey, Abhimanyu and Jauhri, Abhinav and Pandey, Abhinav and Kadian, Abhishek and Al-Dahle, Ahmad and Letman, Aiesha and Mathur, Akhil and Schelten, Alan and Vaughan, Alex and others},
  journal={arXiv preprint arXiv:2407.21783},
  year={2024}
}

@online{chromadb,
  author = {Chroma-core},
  title = {Chroma-core},
  year = {2025},
  url = {https://github.com/chroma-core/chroma},
  urldate = {2025-02-16}
}

@article{crag,
  title={CRAG-comprehensive RAG benchmark},
  author={Yang, Xiao and Sun, Kai and Xin, Hao and Sun, Yushi and Bhalla, Nikita and Chen, Xiangsen and Choudhary, Sajal and Gui, Rongze and Jiang, Ziran and Jiang, Ziyu and others},
  journal={Advances in Neural Information Processing Systems},
  volume={37},
  pages={10470--10490},
  year={2024}
}

@article{RocketKV,
  title={RocketKV: Accelerating Long-Context LLM Inference via Two-Stage KV Cache Compression},
  author={Behnam, Payman and Fu, Yaosheng and Zhao, Ritchie and Tsai, Po-An and Yu, Zhiding and Tumanov, Alexey},
  journal={arXiv preprint arXiv:2502.14051},
  year={2025}
}

@article{xKV,
  title={xkv: Cross-layer svd for kv-cache compression},
  author={Chang, Chi-Chih and Lin, Chien-Yu and Akhauri, Yash and Lin, Wei-Cheng and Wu, Kai-Chiang and Ceze, Luis and Abdelfattah, Mohamed S},
  journal={arXiv preprint arXiv:2503.18893},
  year={2025}
}

@article{triviaqa,
  title={Triviaqa: A large scale distantly supervised challenge dataset for reading comprehension},
  author={Joshi, Mandar and Choi, Eunsol and Weld, Daniel S and Zettlemoyer, Luke},
  journal={arXiv preprint arXiv:1705.03551},
  year={2017}
}

@article{hotpotqa,
  title={HotpotQA: A dataset for diverse, explainable multi-hop question answering},
  author={Yang, Zhilin and Qi, Peng and Zhang, Saizheng and Bengio, Yoshua and Cohen, William W and Salakhutdinov, Ruslan and Manning, Christopher D},
  journal={arXiv preprint arXiv:1809.09600},
  year={2018}
}

@article{2wikimqa,
  title={Constructing a multi-hop qa dataset for comprehensive evaluation of reasoning steps},
  author={Ho, Xanh and Nguyen, Anh-Khoa Duong and Sugawara, Saku and Aizawa, Akiko},
  journal={arXiv preprint arXiv:2011.01060},
  year={2020}
}

@article{googlenq,
  title={Natural questions: a benchmark for question answering research},
  author={Kwiatkowski, Tom and Palomaki, Jennimaria and Redfield, Olivia and Collins, Michael and Parikh, Ankur and Alberti, Chris and Epstein, Danielle and Polosukhin, Illia and Devlin, Jacob and Lee, Kenton and others},
  journal={Transactions of the Association for Computational Linguistics},
  volume={7},
  pages={453--466},
  year={2019},
  publisher={MIT Press One Rogers Street, Cambridge, MA 02142-1209, USA journals-info~…}
}

@article{MV,
author = {Blakeley, Jose A. and Larson, Per-Ake and Tompa, Frank Wm},
title = {Efficiently updating materialized views},
year = {1986},
publisher = {Association for Computing Machinery},
address = {New York, NY, USA},
volume = {15},
number = {2},
journal = {SIGMOD Record},
month = jun,
pages = {61–71}
}

@INPROCEEDINGS{MV2,
  author={Chaudhuri, S. and Krishnamurthy, R. and Potamianos, S. and Shim, K.},
  booktitle={Proceedings of the Eleventh International Conference on Data Engineering (ICDE '95)}, 
  title={Optimizing queries with materialized views}, 
  year={1995},
  pages={190-200}
}

@misc{xiong2024layerkvoptimizinglargelanguage,
      title={LayerKV: Optimizing Large Language Model Serving with Layer-wise KV Cache Management}, 
      author={Yi Xiong and Hao Wu and Changxu Shao and Ziqing Wang and Rui Zhang and Yuhong Guo and Junping Zhao and Ke Zhang and Zhenxuan Pan},
      year={2024},
      eprint={2410.00428},
      archivePrefix={arXiv},
      primaryClass={cs.DC},
      url={https://arxiv.org/abs/2410.00428}, 
}

@inproceedings{InfiniGen,
    author = {Lee, Wonbeom and Lee, Jungi and Seo, Junghwan and Sim, Jaewoong},
    title = {InfiniGen: efficient generative inference of large language models with dynamic KV cache management},
    year = {2025},
    booktitle = {Proceedings of the 18th USENIX Conference on Operating Systems Design and Implementation},
    articleno = {9},
    numpages = {18},
    location = {Santa Clara, CA, USA},
    series = {OSDI'24}
}

@article{InfSurvey,
  title={A survey on efficient inference for large language models},
  author={Zhou, Zixuan and Ning, Xuefei and Hong, Ke and Fu, Tianyu and Xu, Jiaming and Li, Shiyao and Lou, Yuming and Wang, Luning and Yuan, Zhihang and Li, Xiuhong and others},
  journal={arXiv preprint arXiv:2404.14294},
  year={2024},
}

@misc{KVCaching,
    author = {Pierre Lienhart},
    title     = {LLM Inference Series: 3. KV caching explained},
    howpublished = {\url{https://medium.com/@plienhar/llm-inference-series-3-kv-caching-unveiled-048152e461c8}},
    Month = {December},
    year = {2023}
}

@inproceedings{MiniCache,
title={MiniCache: {KV} Cache Compression in Depth Dimension for Large Language Models},
author={Akide Liu and Jing Liu and Zizheng Pan and Yefei He and Gholamreza Haffari and Bohan Zhuang},
booktitle={The Thirty-eighth Annual Conference on Neural Information Processing Systems},
year={2024},
url={https://openreview.net/forum?id=sgVOjDqUMT},
}

@misc{NVIDIA2,
    author = {NVIDIA},
    title     = {NVIDIA H100}, 
    howpublished = {\url{https://www.nvidia.com/en-us/data-center/h100/}},
    Month = {January},
    year = {2025}
}

@inproceedings{FlashAtt,
author = {Dao, Tri and Fu, Daniel Y. and Ermon, Stefano and Rudra, Atri and R\'{e}, Christopher},
title = {FLASHATTENTION: Fast and Memory-Efficient Exact Attention with IO-awareness},
year = {2022},
booktitle = {Proceedings of the 36th International Conference on Neural Information Processing Systems},
articleno = {1189},
numpages = {16},
series = {NIPS '22}
}

@inproceedings{PagedAtt,
author = {Kwon, Woosuk and Li, Zhuohan and Zhuang, Siyuan and Sheng, Ying and Zheng, Lianmin and Yu, Cody Hao and Gonzalez, Joseph and Zhang, Hao and Stoica, Ion},
title = {Efficient Memory Management for Large Language Model Serving with PagedAttention},
year = {2023},
isbn = {9798400702297},
publisher = {Association for Computing Machinery},
address = {New York, NY, USA},
url = {https://doi.org/10.1145/3600006.3613165},
doi = {10.1145/3600006.3613165},
booktitle = {Proceedings of the 29th Symposium on Operating Systems Principles},
pages = {611–626},
numpages = {16},
location = {Koblenz, Germany},
series = {SOSP '23}
}

@article{PromptCache,
  title={Prompt Cache: Modular Attention Reuse for Low-Latency Inference},
  author={In Gim and Guojun Chen and Seung-seob Lee and Nikhil Sarda and Anurag Khandelwal and Lin Zhong},
  journal={ArXiv},
  year={2023},
  volume={abs/2311.04934},
  url={https://api.semanticscholar.org/CorpusID:265067391}
}

@inproceedings{RAG, 
author = {Lewis, Patrick and Perez, Ethan and Piktus, Aleksandra and Petroni, Fabio and Karpukhin, Vladimir and Goyal, Naman and K\"{u}ttler, Heinrich and Lewis, Mike and Yih, Wen-tau and Rockt\"{a}schel, Tim and Riedel, Sebastian and Kiela, Douwe},
title = {Retrieval-augmented generation for knowledge-intensive NLP tasks},
year = {2020},
isbn = {9781713829546},
publisher = {Curran Associates Inc.},
address = {Red Hook, NY, USA},
booktitle = {Proceedings of the 34th International Conference on Neural Information Processing Systems},
articleno = {793},
numpages = {16},
location = {Vancouver, BC, Canada},
series = {NIPS '20}
}

@article{RAGCache,
  title={RAGCache: Efficient Knowledge Caching for Retrieval-Augmented Generation},
  author={Chao Jin and Zili Zhang and Xuanlin Jiang and Fangyue Liu and Xin Liu and Xuanzhe Liu and Xin Jin},
  journal={ArXiv},
  year={2024},
  volume={abs/2404.12457},
  url={https://api.semanticscholar.org/CorpusID:269283058}
}

@misc{9100Pro,
    author = {{Samsung Electronics}},
    title     = {Samsung Announces the 9100 PRO Series SSDs, with Breakthrough PCIe 5.0 Performance}, 
    howpublished = {\url{https://news.samsung.com/us/samsung-announces-9100-pro-series-ssds-with-breakthrough-pcie-5-0-performance/}},
    Month = {Feb},
    year = {2025}
}

@inproceedings{Sarathi,
author = {Agrawal, Amey and Kedia, Nitin and Panwar, Ashish and Mohan, Jayashree and Kwatra, Nipun and Gulavani, Bhargav S. and Tumanov, Alexey and Ramjee, Ramachandran},
title = {Taming throughput-latency tradeoff in LLM inference with sarathi-serve},
year = {2024},
booktitle = {Proceedings of the 18th USENIX Conference on Operating Systems Design and Implementation},
articleno = {7},
numpages = {18},
series = {OSDI'24}
}

@misc{TurboRAG,
      title={TurboRAG: Accelerating Retrieval-Augmented Generation with Precomputed KV Caches for Chunked Text}, 
      author={Songshuo Lu and Hua Wang and Yutian Rong and Zhi Chen and Yaohua Tang},
      year={2024},
      eprint={2410.07590},
      archivePrefix={arXiv},
      primaryClass={cs.CV},
      url={https://arxiv.org/abs/2410.07590}, 
    note={GitHub: \url{https://github.com/MooreThreads/TurboRAG}}
}

@misc{PM9A3,
    author = {{Samsung Electronics}},
    title     = {Samsung SSD PM9A3 (White Paper)}, 
    howpublished = {\url{https://download.semiconductor.samsung.com/resources/white-paper/PM9A3_SSD_Whitepaper.pdf}},
    Month = {May},
    year = {2024}
}

@misc{AgenticAI-inf-size,
  author = {Debmalya Biswas}, 
  title = {Agentic AI Inference Sizing
Understanding Pricing for AI Agents},
  howpublished = {\url{
https://ai.gopubby.com/agentic-ai-inference-sizing-2fd5f9e9578c}}, 
    Month = {March},
    year = {2025}
}

@misc{Text2SQL-prompt-cache,
  author = {Thuwarakesh Murallie},
  title = {Advanced Prompt Caching Techniques to Build Faster \& Cheaper AI},
  howpublished = {\url{https://ai.gopubby.com/advanced-prompt-caching-techniques-to-build-faster-cheaper-ai-391d19a08405}},
    Month = {May},
    year = {2024}
}

@article{lost_in_the_middle, 
year = {2024}, 
title = {Lost in the Middle: How Language Models Use Long Contexts}, 
author = {Liu, Nelson F. and Lin, Kevin and Hewitt, John and Paranjape, Ashwin and Bevilacqua, Michele and Petroni, Fabio and Liang, Percy}, 
journal = {Transactions of the Association for Computational Linguistics}, 
doi = {10.1162/tacl\_a\_00638}, 
pages = {157--173}, 
volume = {12}
}

@article{dejavu,
  title={D$\backslash$'ej$\backslash$aVu: KV-cache Streaming for Fast, Fault-tolerant Generative LLM Serving},
  author={Strati, Foteini and Mcallister, Sara and Phanishayee, Amar and Tarnawski, Jakub and Klimovic, Ana},
  journal={arXiv preprint arXiv:2403.01876},
  year={2024}
}

@article{locomo,
  title={Evaluating very long-term conversational memory of llm agents},
  author={Maharana, Adyasha and Lee, Dong-Ho and Tulyakov, Sergey and Bansal, Mohit and Barbieri, Francesco and Fang, Yuwei},
  journal={arXiv preprint arXiv:2402.17753},
  year={2024}
}

@inproceedings{deep1b,
  title={Efficient indexing of billion-scale datasets of deep descriptors},
  author={Babenko, Artem and Lempitsky, Victor},
  booktitle={Proceedings of the IEEE Conference on Computer Vision and Pattern Recognition},
  pages={2055--2063},
  year={2016}
}

\end{document}